\documentclass[lettersize,journal]{IEEEtran}
\usepackage{amsmath}
\usepackage{cite}
\usepackage{amssymb}
\usepackage{amsfonts}
\usepackage{algorithm}
\usepackage{dsfont}
\usepackage{graphicx}
\usepackage{epsfig}
\usepackage{subfigure}
\usepackage{psfrag}
\usepackage{xcolor}
\usepackage{url}
\usepackage[colorlinks,linkcolor=black,urlcolor=black,anchorcolor=black,citecolor=black,hyperfootnotes=true]{hyperref}
\usepackage{comment}
\usepackage{upgreek}

\newtheorem{remark}{Remark}

\begin{document}

\title{Joint LOS Identification and Data Association for 6G-Enabled Networked Device-Free Sensing}
\author{Qin Shi, \textit{Student Member, IEEE}, and Liang Liu, \textit{Senior Member, IEEE}
\thanks{Manuscript received August 28, 2023; revised December 6, 2023; accepted March 9, 2024. This work was supported in part by the National Key R\&D Project of China under Grant No. 2022YFB2902800; in part by the Research Grants Council, Hong Kong, China, under Grant 25215020 and 15203222. (\it{corresponding author}: Liang Liu)}
\thanks{Qin Shi and Liang Liu are with the Department of Electrical and Electronic Engineering, The Hong Kong Polytechnic University, Hong Kong, China (e-mails: qin-eie.shi@connect.polyu.hk; liang-eie.liu@polyu.edu.hk).}
\thanks{The materials in this paper have been presented in part to the IEEE International Conference on Acoustics, Speech, and Signal Processing (ICASSP) 2023 \cite{shi2023joint}.}}

\maketitle \thispagestyle{empty} \vspace{-30pt}

\begin{abstract}
This paper considers networked device-free sensing in an orthogonal frequency division multiplexing (OFDM) cellular system with multipath environment, where the passive targets reflect the downlink signals to the base stations (BSs) via non-line-of-sight (NLOS) paths and/or line-of-sight (LOS) paths, and the BSs share the sensing information extracted from their received echoes to jointly localize the targets. A two-phase localization protocol is considered. In Phase I, we design an efficient method that is able to accurately estimate the range of any path from a transmitting BS to a receiving BS via a target, even if the transmitting and receiving BSs are separated and not perfectly synchronized. In Phase II, we propose an effective method that is able to jointly identify the ranges of the LOS paths between the targets and the BSs as well as associate the ranges of LOS paths with the right targets, such that the number and the locations of the targets both can be accurately estimated. Numerical results verify that our proposed two-phase protocol can achieve high performance of networked sensing in the multipath environment.

\end{abstract}


\section{Introduction}\label{sec:intro}
\subsection{Background and Motivations}
Recently, integrated sensing and communication (ISAC) has been listed as one of the six key usage scenarios in the future sixth-generation (6G) cellular network by International Telecommunication Union (ITU)\cite{IMT}. Under the ISAC technology, a common radio signal can be used for conveying messages and sensing the environment simultaneously. It is expected that the ISAC technology can play an active role in traffic monitoring at Smart Transportation, fall detection at Smart Hospital, robot tracking at Smart Factory, etc.

Motivated by the significance of the ISAC technology in the 6G network, a lot of attempts for this emerging direction have been made. Many prior works focus on investigating the performance balance between the capacity of communication and the estimation distortion in radar systems\cite{mu2022noma,tsinos2021joint,liu2020joint}, because the optimal waveforms for the communication signals and the sensing signals are quite different \cite{sturm2011waveform}. Apart from the performance optimization, there are also works investigating practical signal processing techniques for embedding the sensing function into the 6G cellular network. For instance, efficient algorithms have been proposed such that a BS can extract the range/angle/Doppler information of the targets based on the OFDM signals \cite{zheng2017super,liu2020two}, the orthogonal time frequency space (OTFS) signals \cite{gaudio2020effectiveness}, and the millimeter wave signals \cite{dokhanchi2019mmwave}, that are reflected by these targets. Moreover, \cite{barneto2022millimeter,yang2022hybrid} have devised powerful estimation schemes such that a mobile user can utilize the cellular signals for realizing simultaneous localization and mapping (SLAM).

It is worth noting that the above works mainly consider the scenario where localization is performed with one transmitter and one collocated/separate receiver, as in the monostatic/bistatic radar systems. Inspired by cooperative communication techniques such as cloud radio access network (C-RAN) \cite{liu2015joint,gesbert2010multi,checko2014cloud}, we are interested in the networked device-free sensing setup \cite{liu2022survey,xie2023collaborative,zhang2020perceptive}, where the targets passively reflect the OFDM signals emitted by the BSs for downlink communication, and the BSs share the local sensing information obtained from their received echoes to jointly localize these targets, as shown in Fig. \ref{system model}. The goal of this paper is to provide practical solutions for incorporating the sensing functions into the cellular network, such that ubiquitous sensing can also be realized in future 6G cellular networks.

\subsection{Prior Works}
There are many interesting and important explorations made for network-level sensing, as discussed in the following.
\subsubsection{Networked Device-based Sensing}
In networked device-based sensing, the targets are able to transmit signals actively such that the mapping between the measurements at each BS and the targets can be known based on the signal signature\cite{liu2022survey}. In this case, if LOS paths can be identified, all BSs can cooperatively localize each target, e.g., time-of-arrival (ToA) based method and angle-of-arrival (AoA) based method \cite{mao2007wireless}. Therefore, it is important to mitigate NLOS paths in networked device-based sensing, which has been well surveyed in various works \cite{aditya2018survey,guvenc2009survey,chen2013comparative}. The simplest way is that the LOS path is assumed to be the shortest and/or the strongest one among all paths, but it may be affected by obstacles in the environment\cite{aditya2018survey}. In \cite{kotaru2015spotfi}, it is observed that the AoA/ToA of the LOS path has a smaller variation over several continuous packets than that of the NLOS path, which is leveraged for identifying the LOS path. The distance bias of NLOS path caused by other objects is modeled as some certain probability
distribution\cite{chen1999non,venkatesh2006linear}. By leveraging the modeled probability distribution, the maximum likelihood (ML) based algorithm is
proposed to estimate the locations of targets. Moreover, the prior knowledge of statistics of LOS path and NLOS path is assumed to be known in \cite{decarli2010nlos}. Then, the likelihood-ratio tests for LOS/NLOS identification are conducted to localize the target.
\subsubsection{Network Device-free Sensing}
When the targets are passive and not equipped with communication functions, device-free sensing is leveraged to estimate their locations. Similarly, the LOS paths between targets and BSs should also be identified for target localization. On the other hand, as pointed out by \cite{shi2022device} where there is no NLOS path or LOS blockage, the data association between ranges of LOS paths and targets also needs to be addressed in networked device-free sensing. Specifically, each BS does not know which (range of) LOS path belongs to which target, since the signatures of signals reflected by all targets are the same. To associate measurements with targets, plenty of pioneering and excellent algorithms have been presented in previous works. One straightforward way is the nearest neighbor (NN) method, which directly associates the most likely measurement to each target. However, the same measurement may be allocated to two or more targets, which is not applicable to the case that one measurement is associated with at most one target, e.g., when any two targets can be distinguished in the range/angle domain. The problem of globally assigning the measurements to the targets can be formulated as a multi-dimensional assignment (MDA) problem\cite{poore1994multidimensional,poore1993lagrangian}. When the dimension of the MDA problem is two, it is a linear assignment problem, which can be solved in polynomial time via the Hungarian algorithm\cite{kuhn1955hungarian}. When the dimension exceeds two, it is an NP-hard problem, which can be solved by branch and bound algorithm and Lagrangian relaxation algorithm \cite{poore1994multidimensional,poore1993lagrangian}. Instead of only retaining the best data association in the MDA method, the PDA and JPDA methods consider each kind of possible data association hypothesis \cite{bar1975tracking, fortmann1983sonar}. Different from considering data association hypotheses at one snapshot, the hypotheses over several consecutive snapshots are jointly taken into account in the MHT method\cite{reid1979algorithm}.

However, these works mainly focus on multiple target tracking (MTT), where prior information about target locations is available for associating measurements with targets at each time slot. With prior target locations as a reference, each BS can individually associate its measurements with these targets. Nevertheless, the locations of targets are unknown in our considered framework, where all BSs need to jointly localize passive targets only based on the obtained measurements. More importantly, LOS identification and data association are coupled together in networked device-free sensing, because each BS does not know either the mapping between paths and targets or that between ranges of LOS paths and targets. In detail, each BS should not only mitigate NLOS paths and identify which BSs have the range information about a particular target, but also perform data association between ranges of identified LOS paths and targets. To our best knowledge, there is a lack of research works that jointly consider the LOS identification and data association problem under networked device-free sensing. Multiple passive target localization in a multi-path environment is studied in \cite{shen2014estimating}, but the mapping constraint between targets and ranges is ignored. Specifically, it is assumed that each range can be associated with multiple targets  due to low range resolution, which will lead to a high probability of the existence of undesired targets. Motivated by this, we propose to jointly tackle the issues of LOS identification and data association under networked device-free sensing, where any two targets can be distinguished in the range domain thanks to the high range resolution provided in the future 6G cellular network.

\begin{figure}
\centering
\includegraphics[width=3.5in]{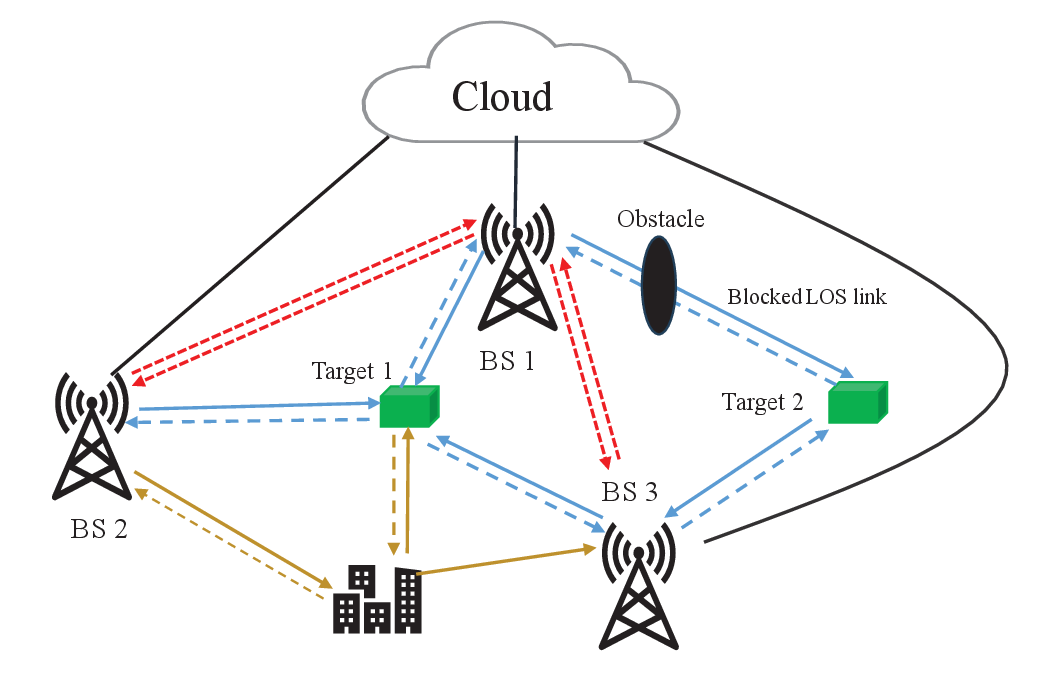}
\caption{System model of our considered networked device-free sensing architecture. The BSs are connected to the central processor via fronthaul links to share the range estimations. For wireless propagation, there are Type I paths (e.g., the path from BS 1 to BS 3), Type II paths (e.g., the path from BS 1 to Target 1 to BS 2), Type III paths (e.g., the path from BS 2 to the building to Target 1
back to BS 3), and LOS blockage (Target 2 cannot be detected by BS 1 due to the obstacle).}
\label{system model}
\end{figure}

\subsection{Main Contributions}
In this paper, we study the networked device-free sensing technique in the OFDM cellular network. As shown in Fig. 1, multiple BSs transmit downlink OFDM signals to convey messages to mobile users. At the same time, each target reflects the OFDM signals to the BSs via the NLOS paths and/or LOS paths. Under this setup, we consider a two-phase protocol to perform networked sensing via BS cooperation. In Phase I, based on its received signals emitted by various BSs, each BS estimates the ranges of all the LOS paths and NLOS paths to it. In Phase II, all the BSs transmit their range information obtained in Phase I to the cloud over the optical fibers, such that the global range information can be utilized to estimate the number and the locations of the targets. Our contributions are summarized as follows.
\begin{itemize}
\item First, in practice, the BSs are not perfectly synchronized. The difference between the clocks at two BSs is defined as the sampling timing offset (STO). For OFDM communication, it is known that as long as the STO is shorter than the length of the cyclic prefix (CP), the effect of STO on communication can be mitigated by removing the CP \cite{schmidl1997robust,morelli2007synchronization}. However, such an approach cannot mitigate the effect of STO on sensing. Specifically, due to the STO, the estimated propagation delay from a transmitting BS to a target to the receiving BS is the sum of the true propagation delay and the STO between the transmitting and the receiving BSs. Therefore, STO will affect the accuracy of range estimation in Phase I. In this paper, we design an efficient method that can estimate both the STO and the effective propagation delay including the STO based on the sparse optimization technique. Therefore, the true propagation delay can be efficiently estimated when the BSs are not perfectly synchronized.
\item In Phase II, we pointed out three challenges to performing target localization based on the range information obtained in Phase I. The first challenge is that the LOS path between a target and a BS may not exist due to the blockage. Therefore, some useful range information is missing at each BS. Besides, due to the multipath environment, a target's echo signal can be received by a BS over the NLOS path. Therefore, some range information obtained by the BSs is not useful. The last one is that at each BS, it does not directly know how to match one estimated range to the right target. This is the data association issue for networked device-free sensing \cite{shi2022device}. Due to the first two challenges, it is difficult to estimate the number of targets based on the cardinality of the range set at each BS. Due to the second and third challenges, we need to jointly perform NLOS mitigation and data association. In other words, for each range estimated in Phase I, we should determine whether it is obtained via a LOS path, and if yes, which target it will belong to.
\item To decouple the above three challenges in Phase II, we propose a novel algorithm. Suppose that the number of BSs is $M$. Under this algorithm, at the first iteration, we aim to estimate the number and locations of the targets that have LOS paths to all $M$ BSs, and in the second iteration, we aim to estimate the number and locations of the targets that have LOS paths just to $M-1$ BSs, and so on. This architecture can tackle the first challenge because at each iteration, we just consider the targets that have LOS paths to a set of the BSs. Next, at each iteration of this algorithm, we propose an efficient method to perform NLOS mitigation and data association to localize the targets that do not have the LOS blockage issue to the considered set of BSs.
\end{itemize}

The rest of this paper is organized as follows. The system model and the sensing signal model are described in Section \ref{sec:model}. Section \ref{sec:Phase I} first introduces how to eliminate the synchronization error between any two BSs. Then, the true delays (ranges) of all paths between the BSs can be  estimated. In Section \ref{sec:Phase II}, an efficient algorithm is proposed to jointly tackle the issues of LOS identification and data association to estimate the number and the locations of targets. The numerical results in Section \ref{sec:numerical results} verify the effectiveness of the proposed algorithm in localizing targets in the multi-path environment. In the end, Section \ref{sec:conclusion} concludes the paper.

\emph{Notations:} Boldface lower-case and boldface upper-case letters are used to represent vectors and matrices, respectively. Denote superscripts ${(\cdot)}^T$ and ${(\cdot)}^H$ by the transpose and the Hermitian operators, respectively. Let $\mathrm{diag}(\cdot)$ and $[\cdot]_{i,j}$ denote the diagonal matrix with the vector being the main diagonal and $(i,j)$-th entry of a matrix, respectively. Denote $\Vert \cdot \Vert_1 $ by the $l_1$ norm of a vector and $|\cdot|$ by the absolute value of a scalar. We use $\tbinom{n}{k}$ for $k$-combinations of the set with $n$ elements, $|\mathcal{A}|$ for the cardinality of the set $\mathcal{A}$, $\mathbb{C}$ for the set of complex values, $\mathbb{C}^{N \times N}$ for the set of complex-value matrices of size $N \times N$, and $\boldsymbol{I}_N$ for the identity matrix of size $N \times N$.

\section{System Model}\label{sec:model}
In this paper, we consider a 6G-enabled ISAC system consisting of $M$ BSs, $K$ targets to be localized (the value of $K$ is unknown and needs to be estimated), and $I$ users for communication, as shown in Fig. \ref{system model}. Since the communication technology is very mature in the cellular network, we mainly focus on the sensing function in this ISAC system. Let $(a_m,b_m)$ and $(x_k,y_k)$ denote the 2D coordinates of the $m$-th BS and the $k$-th target, respectively, $m=1,\ldots,M$, $k=1,\ldots,K$. Then, the direct distance between the $m$-th BS and the $k$-th target is given by
\begin{align}
d_{m,k}&=f(x_k,y_k,a_m,b_m)\nonumber \\ &=\sqrt{(a_m-x_k)^2+(b_m-y_k)^2},~\forall m, k.
\label{eqS2.1}
\end{align}Moreover, the sum of the distance between the $u$-th BS and the $k$-th target and that between the $k$-th target and the $m$-th BS is
\begin{align}
d_{u,m,k}=d_{u,k}+d_{m,k},~\forall u, m, k.
\label{eqS2.2}
\end{align}In the downlink, the BSs will transmit the OFDM signals to the information receivers, while these signals can be reflected by the targets to different BSs as well. In practice, each target may have LOS links to some BSs while having no LOS links to the other BSs, as shown in Fig. \ref{system model}. Define $\mathcal{M}_k$ as the set of BSs that have the LOS links to the $k$-th target.\footnote{In our paper, it is assumed that each target is detected by at least three BSs, i.e., $|\mathcal{M}_k| \geq 3$, $\forall k$.} If there exists a LOS link between target $k$ and BS $m$, we may estimate their distance, i.e., $d_{m,k}$, based on the propagation delay from BS $m$ to target $k$ back to BS $m$. Moreover, if there exist LOS links from target $k$ to both BS $m$ and BS $u\neq m$, then we can also estimate $d_{u,m,k}$ based on the propagation delay from BS $u$ to target $k$ to BS $m$. Last, each target can be localized based on these distance values obtained from the LOS links.

Specifically, let $\boldsymbol{s}_m = [s_{m,1},\ldots,s_{m,N}]^T$ denote one frequency-domain OFDM symbol at the $m$-th BS, $\forall m$, where $s_{m,n}$ is the signal at the $n$-th sub-carrier and $N$ is the number of sub-carriers. Then, the time-domain modulated signal of BS $m$ over one OFDM symbol consisting of $N$ samples is given by $\boldsymbol{\chi}_m = [\chi_{m,1},...,\chi_{m,N}]^T = \sqrt{p}\boldsymbol{W}^H \boldsymbol{s}_m$, $\forall m$, where $p$ denotes the common transmit power for all the BSs, and $\boldsymbol{W} \in \mathbb{C}^{N \times N}$ denotes the discrete Fourier transform (DFT) matrix with $\boldsymbol{W}\boldsymbol{W}^H=\boldsymbol{W}^H\boldsymbol{W}=\boldsymbol{I}$. After inserting the CP consisting of $Q$ OFDM samples, the time-domain signal transmitted by BS $m$ over one OFDM symbol period is given by $\bar{\boldsymbol{\chi}}_{m}=[\bar{\chi}_{m,-Q},\ldots,\bar{\chi}_{m,-1},\bar{\chi}_{m,0},\ldots,\bar{\chi}_{m,N-1}]$, where if $n>0$, $\bar{\chi}_{m,n} = \chi_{m,n+1}$ denotes the useful signal, and if $n\leq 0$, $\bar{\chi}_{m,n} = \chi_{m,N+n+1}$ denotes the CP.

Because the BSs are not perfectly synchronized in practice, define the STO between BS $u$ and BS $m$ as $\tau_{u,m}$ OFDM sample periods\footnote{There are also carrier frequency offset and sampling frequency offset in the network, but we assume that they have been estimated and mitigated by existing methods\cite{schmidl1997robust,morelli2007synchronization}.}, $\forall u,m$, i.e., if the local clock time at BS $m$ is $t_m$, then that at BS $u$ is $t_u=t_m+\tau_{u,m}$. In other words, for BSs $m$ and $u$ that are not perfectly synchronized, $\tau_{u,m}>0$ if the clock time at BS $m$ is earlier, while $\tau_{u,m}<0$ otherwise. Then, define the maximum absolute STO in the network as $\tau_{\rm max}=\text{max}_{u,m}|\tau_{u,m}|$. Moreover, define $\boldsymbol{h}_{u,m}=[h_{u,m,0},\ldots,h_{u,m,L-1}]^T$ as the $L$-tap multi-path channel from BS $u$ to BS $m$, where $h_{u,m,l}$ denotes the complex channel coefficient of the path with a delay of $l$ OFDM sample periods and $L$ denotes the maximum number of resolvable paths. Then, the received time-domain OFDM signal at the $m$-th BS in the $n$-th OFDM sample period, which is contributed by the signals transmitted by the $m$-th BS itself as well as the other $M-1$ BSs that are not perfectly synchronized with the $m$-th BS, can be expressed as
\begin{align}
y_{m,n}&=\sum_{u=1}^M \sum_{l=0}^{L-1} h_{u,m,l} \bar{\chi}_{u,n-l- \tau_{u,m} }+z_{m,n}, ~\forall m,n,
\label{eqS2.4}
\end{align}
where $z_{m,n} \sim \mathcal {CN}(0,\sigma_z^2 )$ denotes the noise at the $m$-th BS in the $n$-th
OFDM sample period. Note that each BS $m$ can potentially receive the signals transmitted by BS $u$ via three types of paths - {\bf Type I path}: the LOS path\footnote{We assume that a Type I path exists between any two BSs because BSs are deployed at high locations such that there is no blockage among them.} from BS $u$ to BS $m$; {\bf Type II path}: the combination of the LOS path from BS $u$ to some target and that from this target to BS $m$; {\bf Type III path}: the NLOS path, e.g., the path from BS $u$ to BS $m$ via not only some target but also some reflector/scatter. Thereby, $h_{u,m,l}\neq 0$ indicates that there exists a Type I/II/III path from BS $u$ to BS $m$, whose propagation delay is of $l$ OFDM sample periods. Specifically, if $h_{u,m,l}\neq 0$ for a particular $l$ is contributed by a Type I path from BS $u$ to BS $m$, then the propagation delay, i.e.,
\begin{align}
l_{u,m}=\big\lfloor \frac{N\Delta f \sqrt{(a_u-a_m)^2+(b_u-b_m)^2}}{c_0} \big\rfloor, ~\forall u,m,
\label{eqS2.4.0}
\end{align} should be of $l_{u,m}=l$ OFDM sample periods, where $c_0$ is the speed of the light and $\lfloor \cdot \rfloor$ denotes the floor function; if $h_{u,m,l}\neq 0$ is contributed by a Type II path from BS $u$ to some target $k$ to BS $m$, then the propagation delay of this path, i.e., $\lfloor d_{u,m,k}/c_0 \rfloor$, is equal to $l$ OFDM sample periods. As will be shown later in this paper, Type I paths are useful for estimating the STOs among the BSs, Type II paths are useful for localizing the targets, while Type III paths are not beneficial to localization and their impact should be mitigated.

Based on the above observation, in this paper, we adopt a two-phase networked device-free sensing protocol. Specifically, in Phase I, each BS $m$ first estimates the channels $h_{u,m,l}$'s based on its received signals, and then sends its delay (thus range) information to a central processor. The main challenge that will be tackled here lies in the unknown STOs $\tau_{u,m}$'s, because if the clocks at different BSs are not perfectly synchronized, the estimated delays will be shifted by STOs. Then, in Phase II, the central processor will estimate the number and the locations of the targets based on the range information sent by all the BSs. Specifically, for each target, we need to first identify the BSs that have the LOS links to this target, and then find the useful LOS range information (corresponding to Type II paths) for localization by mitigating the NLOS range information (corresponding to Type III paths). In the following two sections, we show the details of Phase I and Phase II under the above protocol.

\section{Phase I: Range Estimation}\label{sec:Phase I}
It is challenging to apply the conventional OFDM channel estimation techniques to estimate $h_{u,m,l}$'s based on (\ref{eqS2.4}) because the STOs, i.e., $\tau_{u,m}$'s, are unknown. In this section, we show that even when the BSs are not perfectly synchronized, the channel estimation problem can still be formulated as a sparse signal recovery problem, which can be efficiently solved by the standard compressed sensing techniques, such as LASSO \cite{LASSO}. Then, we will propose an efficient approach that can jointly estimate the STOs and the range of each path based on the estimated channels. 

Specifically, we reformulate the received signals given in (\ref{eqS2.4}) as
\begin{align}
y_{m,n}&\overset{(a)}{=}\sum_{u=1}^M \sum_{l=l_{u,m}}^{L-1} h_{u,m,l} \bar{\chi}_{u,n-l- \tau_{u,m} } \nonumber \\
&=\sum_{u=1}^M \sum_{l=l_{u,m}+\tau_{u,m}}^{L+\tau_{u,m}-1} h_{u,m,l-\tau_{u,m}} \bar{\chi}_{u,n-l} \nonumber \\
&=\sum_{u=1}^M \sum_{l=0}^{L+\tau_{\rm max}-1} \tilde{h}_{u,m,l} \bar{\chi}_{u,n-l }+z_{m,n}
, ~\forall m,n,
\label{eqS2.4.1}
\end{align} where $l_{u,m}$ is the propagation delay for the Type I path between BS $u$ and BS $m$ as given in (\ref{eqS2.4.0}), and $\tilde{h}_{u,m,l}$ is defined as
\begin{align}
\tilde{h}_{u,m,l}=
\left \{
\begin{array}{ll}
h_{u,m,l-\tau_{u,m}},&\text{if}~l \in [l_{u,m}+\tau_{u,m},L+\tau_{u,m}-1],\\
0,&\text{otherwise}.
\end{array}
\right.
\label{eqS2.5}
\end{align}
Therefore, $\tilde{h}_{u,m,l}$ can be interpreted as the virtual channel corresponding to a path from BS $u$ to BS $m$ whose delay is believed by the imperfectly synchronized BSs to be of $l$ OFDM sample periods but actually is of $l-\tau_{u,m}$ OFDM sample periods. In other words, if $\tilde{h}_{u,m,l}\neq 0$, then there is a path from BS $u$ to BS $m$, whose propagation delay is of $l-\tau_{u,m}$ OFDM sample periods. Note that in (\ref{eqS2.4.1}), ($a$) holds because no path with a delay of $l<l_{u,m}$ OFDM sample periods exists between BS $u$ and BS $m$, i.e., $h_{u,m,l} = 0$, $\forall l \in [0,l_{u,m})$. To guarantee that all the inter-symbol interference (ISI) from the last OFDM symbol is received within the CP of the current OFDM symbol for all the BSs, we assume that the length of the CP satisfies $Q>L+\tau_{\rm max}$. Besides, it is assumed that $l_{u,m}+\tau_{u,m} \geq 0$, $\forall u,m$, such that BS $m$ sees no ISI from the next OFDM symbol sent by BS $u$ even when BS $u$'s clock time is earlier than that at BS $m$, i.e., $\tau_{u,m}<0$. Note that the maximum absolute STO of 5G networks is within 130 ns\cite{li2017analysis}. Therefore, as long as the distance between any two BSs is larger than 39 m, which is true in practice, then $l_{u,m}+\tau_{u,m} \geq 0$ will hold, $\forall u,m$.

Because $\tilde{h}_{u,m,l}$'s are defined as (\ref{eqS2.5}), we do not have STOs, i.e., $\tau_{u,m}$'s, in the new received signal model (\ref{eqS2.4.1}), indicating that $\tilde{h}_{u,m,l}$'s can be estimated based on the conventional OFDM channel estimation techniques. However, the cost is the shifted delay estimation for each path - if $\tilde{h}_{u,m,l} \neq 0$, there is a path from BS $u$ to BS $m$ with a delay of $l-\tau_{u,m}$, rather than $l$, OFDM sample periods. If the STOs are unknown, then the delay/range estimation is always incorrect.

Interestingly, we can efficiently tackle the above issue by utilizing the LOS signals between any two BSs. Specifically, among all the paths from BS $u$ to BS $m$, the Type I path, i.e., the LOS path between them, is of the minimum range. Therefore, let us define
\begin{align}
l^{\prime}_{u,m}=\text{min}\{l|\forall l~\text{with}~\tilde{h}_{u,m,l} \neq 0 \},\forall u \neq m.
\label{eqS2.6}
\end{align}
Then, the imperfectly synchronized BSs $u$ and $m$ estimate the delay of their LOS path as $l^{\prime}_{u,m}$ OFDM sample periods. However, we know that the true delay of this LOS path is of $l_{u,m}$ OFDM sample periods, as defined in (\ref{eqS2.4.0}). Therefore, the STOs can be determined as
\begin{align}
\tau_{u,m}=l^{\prime}_{u,m}-l_{u,m},\forall u \neq m.
\label{eqS2.7}
\end{align}
After STOs are known, we can correct the delay estimation made from $\tilde{h}_{u,m,l}$'s. In the following, we show how to achieve the above goals in Phase I of our considered protocol.

According to (\ref{eqS2.4.1}), the frequency-domain signal received at BS $m$ over all the $N$ sub-carriers can be expressed as \cite{hwang2008ofdm,shi2022device}
\begin{align}
\bar{\boldsymbol{y}}_m \!=\! \sqrt{p}\sum_{u=1}^M \mathrm{diag}(\boldsymbol{s}_u) \boldsymbol{G} \tilde{\boldsymbol{h}}_{u,m}\!+ \! \bar{\boldsymbol{z}}_m =\sqrt{p} \tilde{\boldsymbol{G}} \tilde{\boldsymbol{h}}_{m} \!+\! \bar{\boldsymbol{z}}_m, ~ \forall m,
\label{eqS3.1}
\end{align}
where $\tilde{\boldsymbol{h}}_{m}=[\tilde{\boldsymbol{h}}_{1,m},\ldots,\tilde{\boldsymbol{h}}_{M,m}]^T$ with $\tilde{\boldsymbol{h}}_{u,m}=[\tilde{h}_{u,m,0},\ldots,\tilde{h}_{u,m,L+\tau_{max}-1}]^T$, $\boldsymbol{G} \in \mathbb{C}^{N \times (L+\tau_{max})} $ with the $(n,l)$-th element being $G_{n,l} = e^{\frac{-j2\pi (n-1)(l-1)}{N}}$, $\tilde{\boldsymbol{G}}=[\mathrm{diag}(\boldsymbol{s}_1) \boldsymbol{G},\ldots,$ $\mathrm{diag}(\boldsymbol{s}_M) \boldsymbol{G}]$, and $\bar{\boldsymbol{z}}_m= [\bar{z}_{m,1},\ldots,\bar{z}_{m,N}]^T = \boldsymbol{W} \boldsymbol{z}_m \sim \mathcal {CN}(0,\sigma_z^2 \boldsymbol{I})$.

In this paper, we assume that all the BSs know $\boldsymbol{s}_1,\ldots,\boldsymbol{s}_M$ sent by the BSs. For example, in the channel estimation phase for communication, $\boldsymbol{s}_m$'s are pilot signals and can be known by all the BSs. In the data transmission phase, the BSs can exchange the messages $\boldsymbol{s}_m$'s with each other over the fronthaul links as in the cloud radio access network \cite{liu2021uplink}. Hence, $\tilde{\boldsymbol{G}}$ in (\ref{eqS3.1}) is known by all the BSs. Moreover, due to the limited number of targets, reflectors, and scatters, very few elements in $\tilde{\boldsymbol{h}}_{m}$'s are non-zero, i.e., $\tilde{\boldsymbol{h}}_{m}$ is a sparse channel vector, $\forall m$. This motivates us to utilize the LASSO technique to estimate the time-domain channels \cite{LASSO}. Therefore, given any penalty parameter $\alpha>0$, the problem of channel estimation is formulated as
\begin{equation}
\begin{aligned}
& \underset{\tilde{\boldsymbol{h}}_m}{\text{minimize}}
~ \frac{1}{2}\Vert \boldsymbol{\bar{y}}_{m} - \sqrt{p} \tilde{\boldsymbol{G}} \tilde{\boldsymbol{h}}_{m} \Vert_2^2+ \alpha \Vert \tilde{\boldsymbol{h}}_{m} \Vert_1.
\end{aligned}
\label{eqS3.2}
\end{equation}
Note that the above problem is convex and can be solved efficiently by CVX \cite{grant2009cvx}.

Let $\bar{\boldsymbol{h}}_{m}=[\bar{\boldsymbol{h}}_{1,m},\ldots,\bar{\boldsymbol{h}}_{M,m}]^T$ denote the optimal solution to problem (\ref{eqS3.2}), where $\bar{\boldsymbol{h}}_{u,m}=[\bar{h}_{u,m,0},\ldots,\bar{h}_{u,m,L+\tau_{max}-1}]^T$, $u,m=1,\ldots,M$. If $\bar{h}_{u,m,l}\neq 0$ for some $l$, then there exists a path from BS $u$ to BS $m$ whose delay is estimated as $l-\tau_{u,m}$ OFDM sample periods. As shown in the method described in (\ref{eqS2.6}) and (\ref{eqS2.7}), define
\begin{equation}
\bar{l}^{\prime}_{u,m}=\text{min}\{l|\forall l ~ {\rm with} ~ \bar{h}_{u,m,l} \neq 0\},~\forall u \neq m,
\label{eqS3.4}
\end{equation}
as the estimated propagation delay of the Type I path between BS $u$ and BS $m$. Then, the STOs can be estimated as
\begin{equation}
\bar{\tau}_{u,m}=
\left \{
\begin{array}{ll}
\bar{l}^{\prime}_{u,m}-l_{u,m}, &~\text{if}~ u \neq m,\\
0,&~\text{if}~ u=m.
\end{array}
\right.
\label{eqS3.5}
\end{equation}
Once the STOs are estimated, if $\bar{h}_{u,m,l} \neq 0$ with $l\neq \bar{l}^{\prime}_{u,m}$, we claim that there exists a Type II/III path from BS $u$ to BS $m$ with a range of
\begin{equation}
\bar{r}_{u,m,l}= \frac{ (l-\bar{\tau}_{u,m})c_0}{N \Delta f}+\frac{c_0}{2N \Delta f}.
\label{eqS3.6}
\end{equation}
Based on the definitions of Type II and Type III paths in Section \ref{sec:model}, we can know that
\begin{equation}
\bar{r}_{u,m,l}=
\left \{
\begin{array}{ll}
d_{u,m,k_{u,m,l}}+\epsilon_{u,m,k_{u,m,l}},&\text{Type II path}, \\
d_{u,m,k_{u,m,l}}+\epsilon_{u,m,k_{u,m,l}}+\eta_{u,m,l},&\text{Type III path},
\end{array}
\right.
\label{eqS3.7}
\end{equation}
where $k_{u,m,l}$ denotes the index of the target which reflects the signal from BS $u$ to BS $m$ with a delay of $l$ OFDM sample periods, $\epsilon_{u,m,k_{u,m,l}}$ denotes the error caused by the estimation shown in (\ref{eqS3.6}), and $\eta_{u,m,l}$ denotes the bias introduced by the NLOS propagation. Note that under the above estimation rule, the worst-case range estimation error for each target is given by
\begin{equation}
|\bar{d}_{u,m,k_{u,m,l}}-{d}_{u,m,k_{u,m,l}}|\leq \frac{c_0}{2 N \Delta f}\overset{\Delta}{=}\Delta d.
\label{eqS3.7.1}
\end{equation}
For example, in 5G OFDM systems, the channel bandwidth can be up to $B=400$ MHz at the mmWave band according to 3GPP Release 15 \cite{3gpp}. In this case, the worst-case range estimation error is $0.375$ m. Since $\Delta d$ is practically very small, we assume in the sequel that the values of distance for any two Type II paths originated from one BS and reflected back to another BS by two different targets differ by more than $2\Delta d$, thus the corresponding paths are resolvable. Therefore, any two targets will not generate the same ranges of Type II paths in our considered framework.

To summarize, after Phase I of our considered two-phase networked device-free sensing protocol, each BS $m$ will possess $M$ range estimation sets
\begin{align}
\mathcal{D}_{u,m}=\{\bar{r}_{u,m,l}|\forall l ~ {\rm with} ~ \bar{h}_{u,m,l} \neq 0~{\rm and}~ l \neq \bar{l}^{\prime}_{u,m}\},&\nonumber \\
~ u=1,\cdots, M.&
\label{eqS3.8}
\end{align}
Then, each BS $m$ will transmit the above $M$ range sets, i.e., $\mathcal{D}_{1,m}, \ldots, \mathcal{D}_{M,m}$, to the central processor via the fronthaul links. The job of the central processor in Phase II is to first identify the ranges belonging to Type II paths in $\mathcal{D}_{u,m}$'s and then utilize these ranges for estimating the number and the locations of the targets. However, there are some challenges to achieving the above goal. First, for each target, its LOS range information to some BSs may be missing because as shown in Fig. \ref{system model}, in practice, the Type II paths between some targets and some BSs may be blocked. We thus need to identify which BSs have the LOS range information for a particular target to localize it. Second, even if a target's LOS ranges corresponding to Type II paths are known to be contained in some $\mathcal{D}_{u,m}$'s, they are mixed with many NLOS ranges corresponding to Type III paths. We thus need to mitigate the effects of NLOS propagation for localization. Last, even if $\bar{r}_{u,m,l}$ is identified to be the range estimation associated with a Type II path from BS $u$ to BS $m$ via some target, we do not know whether $\bar{r}_{u,m,l}$ is an estimation of $d_{u,m,1}$ for target 1, $\ldots$, or $d_{u,m,K}$ for target $K$. We thus need to perform data association such that each useful range of a Type II path can be matched to the right target for localizing it\cite{shi2022device}. In the next section, we will show how the central processor can tackle the above challenges in Phase II to estimate the number and the locations of the targets based on $\mathcal{D}_{u,m}$'s.

\section{Phase II: Target Number and Location Estimation}\label{sec:Phase II}
With the range sets $\mathcal{D}_{u,m}$'s, $\forall u,m$, the objectives of the central processor in Phase II are two-fold. First, it needs to estimate the number of targets in the network, i.e., $K$. Second, it needs to estimate the coordinates of the $K$ targets, i.e., $(x_k,y_k)$, $k=1,\ldots,K$. To achieve the above goals, in this section, we will first formulate a problem for joint target number and location estimation. Then, we will propose an efficient LOS identification and data association algorithm to solve the above problem.

For convenience, given any set $\mathcal{D}$, let $\mathcal{D}(g)$ denote its $g$-th smallest element. If the Type II LOS path from BS $u$ to target $k$ and to BS $m$ exists and the estimation of $d_{u,m,k}$ is contained in the set $\mathcal{D}_{u,m}$, then define $g_{u,m,k}>0$ as an integer such that $\mathcal{D}_{u,m}(g_{u,m,k})$ is the estimation of $d_{u,m,k}$. Otherwise, define $g_{u,m,k}=0$. Note that $g_{m,m,k}>0$ and $g_{u,u,k}>0$ is equivalent to $g_{u,m,k}>0$ and vice versa. Moreover, define
\begin{align}
\mathcal{G}_k=\{g_{u,m,k}, \forall u,m\}
\label{eqS4.0}
\end{align} as the solution of LOS identification and data association for target $k$ to all the $M$ BSs, $k=1,\ldots, K$. This is because (i) among all the elements in $\mathcal{D}_{u,m}$, only $\mathcal{D}_{u,m}(g_{u,m,k})$'s, $\forall k$ with $g_{u,m,k}>0$, are the estimated ranges for Type II LOS paths; (ii) if $g_{u,m,k}>0$, then $\mathcal{D}_{u,m}(g_{u,m,k})$ is the estimated range of a Type II path for target $k$. Therefore, if $\mathcal{G}_k$'s are known, we can know which BSs have the LOS links to target $k$, i.e., the set of BSs that can detect target $k$ is given by
\begin{align}
\mathcal{M}_k=\mathcal{M}(\mathcal{G}_k)=\{m|g_{m,m,k}>0, \forall g_{m,m,k} \in \mathcal{G}_k\}.
\label{eqS4.1}
\end{align}
Furthermore, we can find the range estimations belonging to Type II LOS paths from $\mathcal{D}_{u,m}$'s and associate them with the right targets.

For convenience, define $\bar{\mathcal{G}}_k$ as the set containing all the positive integer elements in $\mathcal{G}_k$, $\forall k$. If $\mathcal{G}_1,\ldots,\mathcal{G}_K$ can be found out, then the number of the targets can be directly known because one mapping corresponds to one target and vice versa. Further, given $\mathcal{G}_k$, the location of each target $k$ can be estimated based on its range information $\mathcal{D}_{u,m}(g_{u,m,k})$'s, $\forall u,m$, i.e., trilateration or multilateration localization. In the following, we show how to estimate the number of the targets and their locations via solving the mapping solution $\mathcal{G}_1,\ldots,\mathcal{G}_K$.

\subsection{Problem Formulation}
First, we define the conditions that a feasible mapping solution of $\mathcal{G}_1,\ldots,\mathcal{G}_K$ should satisfy. Given $u,m$, the number of elements in the range set $\mathcal{D}_{u,m}$ is denoted by $N_{u,m}=|\mathcal{D}_{u,m}|$. Since the range associated with each target must originate from the range sets, the elements in $\mathcal{G}_1,\ldots,\mathcal{G}_K$ should satisfy
\begin{align}
g_{u,m,k}\in \{0,1,\ldots,N_{u,m}\}, ~ \forall u,m,k.
\label{eqS4.2}
\end{align}Moreover, if one range estimation in $\mathcal{D}_{u,m}$ is matched to target $k$, then it cannot be matched to another user $\bar{k}\neq k$, i.e.,
\begin{align}
g_{u,m,k}\neq g_{u,m,\bar{k}}, ~ \forall g_{u,m,k} \in \bar{\mathcal{G}}_k, g_{u,m,\bar{k}} \in \bar{\mathcal{G}}_{\bar{k}}.
\label{eqS4.3}
\end{align}
Another condition that $\mathcal{G}_1,\ldots,\mathcal{G}_K$ should satisfy arises from (\ref{eqS2.2}): the length of the Type II path from BS $u$ to target $k$ to BS $m$, i.e., $d_{u,m,k}$, is equal to the sum of the distance between BS $u$ and target $k$, i.e., $d_{u,k}$, and that between BS $m$ and target $k$, i.e., $d_{m,k}$. Note that the imperfect estimations of $d_{u,m,k}$, $d_{u,k}$, and $d_{m,k}$'s are $\mathcal{D}_{u,m}(g_{u,m,k})$ (also $\mathcal{D}_{m,u}(g_{m,u,k})$), $\mathcal{D}_{u,u}(g_{u,u,k})/2$, and $\mathcal{D}_{m,m}(g_{m,m,k})/2$. Therefore, we set the following constraints for $\mathcal{G}_1,\ldots,\mathcal{G}_K$:
\begin{align}
\left|\frac{\mathcal{D}_{u,u}(g_{u,u,k})}{2}\!+\!\frac{\mathcal{D}_{m,m}(g_{m,m,k})}{2}\!-\!\mathcal{D}_{u,m}(g_{u,m,k})\right| \leq \delta,&
\label{eqS4.4} \\
\left|\frac{\mathcal{D}_{u,u}(g_{u,u,k})}{2}\!+\!\frac{\mathcal{D}_{m,m}(g_{m,m,k})}{2}\!-\! \mathcal{D}_{m,u}(g_{m,u,k})\right| \leq \delta, &\nonumber \\
\forall g_{u,m,k} \in \bar{\mathcal{G}}_k,&
\label{eqS4.5}
\end{align}where $\delta>0$ is a given threshold. The last constraint about $\mathcal{G}_1,\ldots,\mathcal{G}_K$ is on the localization residual associated with this solution of LOS identification and data association. Specifically, given $\mathcal{G}_k$, the location of target $k$ is estimated by solving the following nonlinear least squared (NLS) problem
\begin{align*}
{\rm (P1)} ~ \mathop{\textup{minimize}}_{x_k,y_k}~ \sum\limits_{(u,m):g_{u,m,k} \in \bar{\mathcal{G}}_k} (f(x_k,y_k,a_u,b_u)&\nonumber \\ +f(x_k,y_k,a_m,b_m) -\mathcal{D}_{u,m}(g_{u,m,k}))^2,~\forall k,&
\end{align*}where $f(x_k,y_k,a_m,b_m)$'s are given in (\ref{eqS2.1}). Problem (P1) is a non-convex problem. We can adopt the Gauss-Newton method to solve it \cite{torrieri1984statistical}. Given $\mathcal{G}_k$ for target $k$, define $R(\mathcal{G}_k)$ as the value of problem (P1) achieved by the Gauss-Newton method. Therefore, $R(\mathcal{G}_k)$ can be interpreted as the residual for localizing target $k$ given $\mathcal{G}_k$. If $\mathcal{G}_k$ is the right solution, then the localization residual $R(\mathcal{G}_k)$ should be small, $\forall k$, because the correct ranges are associated with each target. For example, when the range estimation error is zero, the localization residual will also be zero if the correct ranges are utilized for localizing each target. We thus set the following residual constraints about $\mathcal{G}_k$'s:
\begin{align}
R(\mathcal{G}_k)\leq \beta, ~ k=1,\ldots,K, \label{eqS4.6}
\end{align}where $\beta>0$ is some given threshold.

To summarize, any $\mathcal{G}_1,\ldots,\mathcal{G}_K$ satisfying constraints (\ref{eqS4.2})-(\ref{eqS4.6}) can be a feasible solution for LOS identification and data association. Hence, we need to decide which one is optimal among all feasible mapping solutions. When there are $K$ targets, it is not likely that the range data estimated from the NLOS paths at various BSs can be well matched such that more than $K$ targets can be detected. In other words, the probability that the number of mappings satisfying constraints (\ref{eqS4.2})-(\ref{eqS4.6}) is larger than that of the true targets (mappings) is low\cite{neira2001data,bailey2000data,grimson1991object}. Therefore, we want to maximize the number of targets whose locations can be estimated in this paper. Specifically, the solution of LOS identification and data association can be found by solving the following problem
\begin{align}
({\rm P2})~\mathop{\textup{maximize}}_{K,\mathcal{G}_1,\ldots,\mathcal{G}_K}&~K \label{eqS4.7}\\
\textup{subject to}&~(\ref{eqS4.2})-(\ref{eqS4.6}). \nonumber
\end{align}
Next, we focus on solving problem (P2) to estimate the number of targets and the mapping solutions. With the obtained mapping solution, the location of each target can be estimated by solving problem (P1).

\subsection{The Proposed Algorithm}
There are three challenges to solve problem (P2). The first challenge is that a target may have no LOS paths to some BSs. Therefore, the range information of many Type II paths is missing in the range set of each BS, i.e., there are $K_{u,m} \leq K$ targets detected by BS $u$ and BS $m$, $\forall u,m$. The number of possibilities for assigning the $N_{u,m}$ ranges in $\mathcal{D}_{u,m}$ to the detected $K_{u,m}$ out of $K$ targets is $\tbinom{N_{u,m}}{K_{u,m}}\tbinom{K}{K_{u,m}}(K_{u,m}!)$. When there is no LOS blockage, i.e., $K_{u,m}=K$, $\forall u,m$, there are $\tbinom{N_{u,m}}{K}(K!)$ assignment possbilities. However, when there is LOS blockage, i.e., $K_{u,m}$ ranges from 0 to $K$, the number of possibilities is $\sum_{K_{u,m}=0}^{K} \tbinom{N_{u,m}}{K_{u,m}}\tbinom{K}{K_{u,m}}(K_{u,m}!)$, which is much larger than the case without LOS blockage. The second one is that a target may reflect the signals to the BSs via NLOS paths. Due to the Type III paths, the cardinality of the range set at each BS is larger, i.e., $N_{u,m}$ is larger, such that the number of assignment probabilities for the range set $\mathcal{D}_{u,m}$ also increases. Furthermore, we have to mitigate the range information of the Type III paths, which are not beneficial for target localization. The last one is that given the range of a Type II path, how to match it to the corresponding target is also a big challenge. Moreover, the above three challenges are coupled together. To localize each target, we need to jointly determine which BSs have LOS paths to it, and for these BSs, which ranges in the range sets belong to this target. In the following, we propose an iterative algorithm to decouple Challenge 1 from Challenges 2 and 3. Specifically, the targets can be classified into the following categories: the targets that have LOS paths to exactly M BSs, the targets that have LOS paths to exactly $M-1$ BSs, and so on. The last category will be the targets that have LOS paths to exactly 3 BSs\footnote{We do not consider the targets that have LOS paths to only one or two BSs because they cannot be localized based on the range information.}.

Define the number of targets that have LOS paths to exactly $l$ BSs as $K_l$, $l=3,\ldots,M$. Then, we have
\begin{align}
K=\sum_{l=3}^M K_l.
\label{eqS4.9}
\end{align}
Under our proposed algorithm, at the first iteration, we aim to localize all the targets that have LOS paths to exactly $M$ BSs, while at the second iteration, we aim to localize all the targets that have LOS paths to exactly $M-1$ BSs, and so on. Note that at each iteration of our proposed algorithm, we do not face Challenge 1, because when we localize some targets, we only consider the BSs that have LOS paths to them. Specifically, the problem of localizing the targets that have LOS paths to $l$ BSs can be formulated as
\begin{align}
({\rm P3-}l)~\mathop{\textup{maximize}}_{K_l,\mathcal{G}_{1}^{(l)},\ldots,\mathcal{G}_{K_l}^{(l)}}&~ K_l \label{eqS4.10} \\
\textup{subject to}~~~& |\mathcal{M}(\mathcal{G}_{k}^{(l)})|=l, ~k=1,\ldots,K_l, \label{eqS5.1.1} \\
~~&(\ref{eqS4.2})-(\ref{eqS4.6}). \nonumber
\end{align}
where $\mathcal{G}_k^{(l)}$ denotes the mapping for target $k$ that can be detected by $l$ BSs and $\mathcal{M}(\mathcal{G}_{k}^{(l)})$ is given in (\ref{eqS4.1}). In the following, we first focus on how to deal with problem (P3$-l$), $\forall l$. Then, based on (\ref{eqS4.9}), the solutions to problem (P3$-l$), $l=M,\ldots,3$, are used for solving problem (P2) such that the number of targets and the mapping solutions can be estimated. Last, with the obtained mapping solution, the location of each target can be estimated by solving problem (P1).

Problem (P3$-l$) can be solved by exhaustive search, i.e., given each $K_l$ and $\mathcal{G}_{1}^{(l)},\ldots,\mathcal{G}_{K_l}^{(l)}$, we check whether conditions (\ref{eqS4.2})-(\ref{eqS4.6}) and (\ref{eqS5.1.1}) hold. However, such an approach needs to solve the non-convex problem (P1) many times, which is of high complexity. To resolve this issue, we first ignore constraints (\ref{eqS4.3}) and (\ref{eqS4.6}) in problem (P3$-l$), and find the feasible region that any mapping $\mathcal{G}_k$ should satisfy constraints (\ref{eqS4.2}), (\ref{eqS4.4}), (\ref{eqS4.5}), and (\ref{eqS5.1.1}), i.e.,
\begin{align}
\bar{\mathcal{G}}^{(l)}=\{\mathcal{G}_k^{(l)}|\mathcal{G}_k^{(l)} \in {\mathcal{G}}^{(l)} ~\text{satisfies}~ (\ref{eqS4.4})~\text{and}~(\ref{eqS4.5}) \}.
\label{eqS5.2}
\end{align}
where 
\begin{align}
{\mathcal{G}}^{(l)}=\{\mathcal{G}_k^{(l)}|\mathcal{G}_k^{(l)}~\text{satisfies}~ (\ref{eqS4.2}) ~\text{and}~ (\ref{eqS5.1.1})\}.
\label{eqS5.2.1}
\end{align}
Due to the constraints (\ref{eqS4.4}) and (\ref{eqS4.5}), the size of $\bar{\mathcal{G}}^{(l)}$ is generally small, because the probability that (\ref{eqS4.4}) and (\ref{eqS4.5}) holds for Type III paths is very small. Then, we just need to check which mappings in $\bar{\mathcal{G}}^{(l)}$ rather than ${\mathcal{G}}^{(l)}$ satisfy the constraint (\ref{eqS4.6}). As a result, the number of times to solve problem (P1) is significantly reduced compared to the exhaustive search approach to problem (P3$-l$). Define
\begin{align}
\tilde{\mathcal{G}}^{(l)}=\{\mathcal{G}_k^{(l)}|\mathcal{G}_k^{(l)}\in \bar{\mathcal{G}}^{(l)} ~ {\rm and} ~ R(\mathcal{G}_k^{(l)})\leq \beta\}.
\label{eqS5.3}
\end{align}
In other words, by removing $\mathcal{G}_k^{(l)}$'s that do not satisfy condition (\ref{eqS4.6}) from $\bar{\mathcal{G}}^{(l)} $, we can obtain $\tilde{\mathcal{G}}^{(l)}$. Note that each element contained in $\tilde{\mathcal{G}}^{(l)}$ satisfies conditions (\ref{eqS4.2}), (\ref{eqS4.4})-(\ref{eqS4.6}), i.e., it is a feasible solution of LOS identification and data association to localize one target. However, for two mappings $\mathcal{G}_k^{(l)}\in \tilde{\mathcal{G}}^{(l)}$ and $\mathcal{G}_{\bar{k}}^{(l)}\in \tilde{\mathcal{G}}^{(l)}$, it is possible that $g_{u,m,k} \in \mathcal{G}_k^{(l)}$ is equal to $g_{u,m,\bar{k}} \in \mathcal{G}_{\bar{k}}^{(l)}$ for some $u,m$, i.e., in these two solutions, some estimated range at BS $m$ is matched to both target $k$ and target $\bar{k}$. Therefore, the last step to solve problem (P3$-l$) is to select the maximum number of elements in $\tilde{\mathcal{G}}$ such that condition (\ref{eqS4.3}) can be satisfied. Such a problem can be formulated as
\begin{align}
({\rm P4})~\mathop{\textup{maximize}}_{K_l,\mathcal{G}_1^{(l)},\ldots,\mathcal{G}_K^{(l)}}&~K_l \label{eqS5.4}\\
\textup{subject to} &~ \mathcal{G}_k^{(l)} \in \tilde{\mathcal{G}}^{(l)}, k=1,\ldots,K_l, \label{eqS5.4.1} \\
&~ (\ref{eqS4.3}). \nonumber
\end{align}

One way to solve problem (P4) is the exhaustive search method. For a given $K_l$, we need try $\tbinom{N_{\tilde{\mathcal{G}}^{(l)}}}{K_l}$ possiblities to see whether (\ref{eqS4.3}) is satisfied, where $N_{\tilde{\mathcal{G}}^{(l)}}$ denotes the cardinality of $\tilde{\mathcal{G}}^{(l)}$. Because we do not know the exact value of $K_l$, we need to try each possible value of $K_l$, which is quite computationally prohibitive. To alleviate the complexity, we can divide $\tilde{\mathcal{G}}^{(l)}$ into $N_{1,1}+1$ subsets based on the value of $g_{1,1,k}$ , i.e.,
\begin{align}
\tilde{\mathcal{G}}^{(l)}=\tilde{\mathcal{G}}_0^{(l)} \cup\tilde{\mathcal{G}}_1^{(l)} \cup \ldots \cup \tilde{\mathcal{G}}_{N_{1,1}}^{(l)},
\label{eqS5.5}
\end{align}
where
\begin{align}
\tilde{\mathcal{G}}_i^{(l)}=\{\mathcal{G}_k^{(l)}| \mathcal{G}_k^{(l)} \in \tilde{\mathcal{G}}^{(l)}~\text{and}~g_{1,1,k}=i~\text{with}~g_{1,1,k} \in \mathcal{G}_k^{(l)} \}& \nonumber \\
i=0,1,\ldots,N_{1,1}.&
\label{eqS5.6}
\end{align} As a consequence, at most one mapping $\mathcal{G}_k^{(l)}$ can be selected from each $\tilde{\mathcal{G}}_i^{(l)}$ to make sure (\ref{eqS4.3}) is satisfied. In other words, some $\tilde{\mathcal{G}}_i^{(l)}$'s contain the mappings $\mathcal{G}_k^{(l)}$'s, while others do not. Note that the number of $\tilde{\mathcal{G}}_i^{(l)}$'s containing the mappings $\mathcal{G}_k^{(l)}$'s is equal to $K_l$, i.e., the $K_l$ constraints in (\ref{eqS5.4.1}). Motivated by this, we aim at finding the indexes of $\tilde{\mathcal{G}}_i^{(l)}$'s that contain the mappings $\mathcal{G}_k^{(l)}$'s. As a result, problem (P4) can be transformed into
\begin{align}
({\rm P5})~\mathop{\textup{maximize}}_{\Omega,\mathcal{G}_1^{(l)},\ldots,\mathcal{G}_{|\Omega|}^{(l)}}&~|\Omega| \label{eqS5.7}\\
\textup{subject to} ~&~ \Omega \subseteq \{0,1,\ldots,N_{1,1}\}, \label{eqS5.8} \\
&~ \mathcal{G}_k^{(l)} \in \tilde{\mathcal{G}}_{\Omega(k)}^{(l)},~k=1,\ldots,|\Omega|, \label{eqS5.9} \\
&~ (\ref{eqS4.3}). \nonumber
\end{align}
Given $\Omega$ in problem (P5), there may be multiple feasible mapping solutions that satisfy (\ref{eqS5.9}) and (\ref{eqS4.3}). Compared with problem (P4), problem (P5) is of lower complexity, since the cases that two or more mappings $\mathcal{G}_k^{(l)}$'s belong to the same $\tilde{\mathcal{G}}_i^{(l)}$ need not be considered. However, it is still of high complexity for large $N_{\tilde{\mathcal{G}}^{(l)}}$. To tackle this issue, we greedily maximize $|\Omega|$ iteratively. At the beginning, we initialize $\Omega^{(0)}=\{\}$. At the $t$-th iteration, we try to increase $|\Omega|$ by 1. Specifically, we set $\Omega=\Omega^{(t-1)}\cup \{t-1\}$ in problem (P5) to see whether there exists a mapping solution $\mathcal{G}_1^{(l)},\ldots,\mathcal{G}_{|\Omega|}^{(l)}$ that satisfies (\ref{eqS4.3}) and (\ref{eqS5.9}). If such a mapping solution do exist, we update $\Omega^{(t)}=\Omega^{(t-1)}\cup \{t-1\}$; otherwise, $\Omega^{(t)}=\Omega^{(t-1)}$. By repeating the above procedure until $t>N_{1,1}+1$, we can get the solution of problem (P5), denoted by $\hat{\Omega}=\Omega^{(N_{1,1}+1)}$.

\begin{algorithm}[t]
{\bf Input}: $(a_m,b_m)$'s, $\mathcal{D}_{u,m}$'s, $\forall u,m$, and $l$ \\
{\bf Procedure}:
\begin{enumerate}
\item[1.] Find $\tilde{\mathcal{G}}^{(l)}$ based on (\ref{eqS5.3}) and get $\tilde{\mathcal{G}}_i^{(l)}$'s according to (\ref{eqS5.6});
\item[2.] Initialize $t=1$ and $\Omega^{(0)}=\{\}$; Get $\boldsymbol{\mathcal{G}}_{1}^{(0)}=\{\}$; Set $k=1$;
\item[3.] {\bf Repeat}:
\begin{enumerate}
\item[3.1.] For each $\boldsymbol{\mathcal{G}}_{n_{t-1}}^{(t-1)}$, check whether there exists $ \mathcal{G}_k^{(l)} \in \tilde{\mathcal{G}}_{t-1}^{(l)}$ that satisfies (\ref{eqS4.3}) with any mapping in $\boldsymbol{\mathcal{G}}_{n_{t-1}}^{(t-1)}$;
\item[3.2.] If such $\mathcal{G}_k^{(l)}$ and $\boldsymbol{\mathcal{G}}_{n_{t-1}}^{(t-1)}$ exist or $\boldsymbol{\mathcal{G}}_{n_{t-1}}^{(t-1)}$ is empty, add $\boldsymbol{\mathcal{G}}_{n_t}^{(t)}=\boldsymbol{\mathcal{G}}_{n_{t-1}}^{(t-1)}\cup \{\mathcal{G}_k^{(l)}\}$ as the mapping solution for $\Omega=\Omega^{(t-1)}\cup \{ t-1\}$. After finding all such $\boldsymbol{\mathcal{G}}_{n_t}^{(t)}$'s, update $\Omega^{(t)}=\Omega^{(t-1)} \cup \{ t-1\}$ and $k=k+1$; otherwise, only update $\Omega^{(t)}=\Omega^{(t-1)}$ and $\boldsymbol{\mathcal{G}}_{n_t}^{(t)}=\boldsymbol{\mathcal{G}}_{n_{t-1}}^{(t-1)}$ for each $\boldsymbol{\mathcal{G}}_{n_{t-1}}^{(t-1)}$;
\item[3.3.] Update $t=t+1$;
\end{enumerate}
{\bf Until} $t>N_{1,1}+1$.
\item [4.] Get the index solution to problem (P5), denoted by $\hat{\Omega}=\Omega^{(N_{1,1}+1)} $. If there are multiple mapping solutions, i.e., $\boldsymbol{\mathcal{G}}_{n_{N_{1,1}+1}}^{(N_{1,1}+1)}$'s, $n_{N_{1,1}+1}=1,2,\ldots,$ select the one minimizing the localization residual as the optimal mapping solution, denoted by $(\boldsymbol{\mathcal{G}}^{(l)})^{\ast}$.
\end{enumerate}
{\bf Output}:

1) Obtain the estimated number and the solution of LOS identification and data association for the targets that are detected by $l$ BSs, denoted by $\hat{K}^{(l)}=|\hat{\Omega}|$ and $\{\hat{\mathcal{G}}_k^{(l)}\}_{k=1}^{\hat{K}}=(\boldsymbol{\mathcal{G}}^{(l)})^{\ast}$, respectively;\\
2) Estimate the target locations by solving problem ({\rm P1}) given the mapping solution $\hat{\mathcal{G}}_k^{(l)}$'s, denoted by $(\hat{x}_k,\hat{y}_k)$'s, $k=1,\ldots,\hat{K}$.

\caption{Algorithm for Solving Problem (P3$-l$)}
\label{alg:1}
\end{algorithm}

Note that if $\mathcal{G}_{1}^{(l)},\ldots, \mathcal{G}_{|\Omega^{(t)}|}^{(l)}$ is a mapping solution for $\Omega=\Omega^{(t)}=\Omega^{(t-1)}\cup \{t-1\}$ in problem (P5), then $\mathcal{G}_{1}^{(l)},\ldots, \mathcal{G}_{|\Omega^{(t-1)}|}^{(l)}$ is a solution for $\Omega=\Omega^{(t-1)}$. This can be utilize to simplify finding $\mathcal{G}_1^{(l)},\ldots,\mathcal{G}_{|\Omega|}^{(l)}$ at the $t$-th iteration. Denote $\boldsymbol{\mathcal{G}}_{n_t}^{(t)}=\{ \mathcal{G}_{n_k}^{(l)}\}_{k=1}^{|\Omega^{(t)}|}$ by the $n_t$-th mapping solution for $\Omega=\Omega^{(t)}$ in problem (P5) at the $t$-th iteration. Then, we just need to check whether there exists $\mathcal{G}_k^{(l)} \in \tilde{\mathcal{G}}_{t-1}^{(l)}$ that satisfy (\ref{eqS4.3}) with any mapping element in $\boldsymbol{\mathcal{G}}_{n_{t-1}}^{(t-1)}$ at the $t$-th iteration. If such $\mathcal{G}_k^{(l)}$ and $\boldsymbol{\mathcal{G}}_{n_{t-1}}^{(t-1)}$ exist or $\boldsymbol{\mathcal{G}}_{n_{t-1}}^{(t-1)}$ is empty, we can update $\Omega^{(t)}=\Omega^{(t-1)} \cup \{t-1\}$ and add $\boldsymbol{\mathcal{G}}_{n_t}^{(t)}= \boldsymbol{\mathcal{G}}_{n_{t-1}}^{(t-1)} \cup \{\mathcal{G}_k^{(l)} \}$ as a feasible mapping solution for $\Omega=\Omega^{(t)}$; otherwise, we keep $\Omega^{(t)}=\Omega^{(t-1)}$ and update $\boldsymbol{\mathcal{G}}_{n_t}^{(t)}= \boldsymbol{\mathcal{G}}_{n_{t-1}}^{(t-1)}$ for each $\boldsymbol{\mathcal{G}}_{n_{t-1}}^{(t-1)}$ as the mapping solution for $\Omega=\Omega^{(t)}$.

The details of the above process are shown in Algorithm \ref{alg:1}. Since there may be multiple mapping solutions, we will try each of them in Step 3.1. If some $\boldsymbol{\mathcal{G}}_{n_{t-1}}^{(t-1)}$'s can lead to $\Omega^{(t)}=\Omega^{(t-1)} \cup \{ t\}$ while others cannot, we only update $\boldsymbol{\mathcal{G}}_{n_t}^{(t)}=\boldsymbol{\mathcal{G}}_{n_{t-1}}^{(t-1)}\cup \{\mathcal{G}_k^{(l)}\}$ for these $\boldsymbol{\mathcal{G}}_{n_{t-1}}^{(t-1)}$'s as the mapping solutions for $\Omega=\Omega^{(t)}=\Omega^{(t-1)} \cup \{t-1\}$. If no $\boldsymbol{\mathcal{G}}_{n_{t-1}}^{(t-1)}$ can lead to $\Omega^{(t)}=\Omega^{(t-1)} \cup \{ t-1\}$ , then we keep $\boldsymbol{\mathcal{G}}_{n_t}^{(t)}=\boldsymbol{\mathcal{G}}_{n_{t-1}}^{(t-1)}$ for each $\boldsymbol{\mathcal{G}}_{n_{t-1}}^{(t-1)}$ as the mapping solution for $\Omega=\Omega^{(t)}=\Omega^{(t-1)}$. In the end, there may be multiple mapping solutions for $\Omega=\hat{\Omega}$ in problem (P5). For the $n$-th mapping solution, denoted by $\{\hat{\mathcal{G}}_{n_k}^{(l)}\}_{k=1}^{|\hat{\Omega}|}$, its localization residual, i.e., $\sum_{k=1}^{|\hat{\Omega}|} \mathcal{R}(\hat{\mathcal{G}}_{n_k}^{(l)})$, is calculated. Then, we select the one with the smallest localization residual as the optimal mapping solution for the estimated $|\hat{\Omega}|$ targets, denoted by $\{ \hat{\mathcal{G}}_k^{(l)}\}_{k=1}^{|\hat{\Omega}|}$.

Next, we show how to get the solution problem (P2) by iteratively solving problem (P3$-l$). We sequentially solve problem (P3$-l$) for $l$ from $M$ to 3 and obtain the mapping solutions for each $l$, denoted by $\{\hat{\mathcal{G}}_k^{(l)}\}_{k=1}^{\hat{K}_l}$. In this process, the ranges corresponding to $\{\hat{\mathcal{G}}_k^{(l+1)}\}_{k=1}^{\hat{K}_{l+1}}$ are removed from the range sets before solving problem (P3-$l$), $\forall l$. As a result, we can find the targets that are exactly detected by $l$ BSs when solving problem (P3$-l$) for each $l$. Specifically, at the $t$-th step, $t=1,\ldots,M-2$, we find the mapping solution for the targets that can be detected by $M_t=M-t+1$ BSs based on Algorithm \ref{alg:1}. Denote the obtained solution at the $t$-th step by $\hat{K}_{M_t}$ and $\hat{\mathcal{G}}_1^{(M_t)},\ldots,\hat{\mathcal{G}}_{K_{M_t}}^{(M_t)}$, which are regarded as the estimated number and the mapping solution for the targets that are exactly detected by $M_t$ BSs. Then, we remove these assigned ranges, $\mathcal{D}_{u,m}(\hat{g}_{u,m,k})$'s, $\forall \hat{g}_{u,m,k} \in \hat{\mathcal{G}}_k^{(M_t)}$ with $\hat{g}_{u,m,k}>0$, $k=1,\ldots,\hat{K}_{M_t}$, from $\mathcal{D}_{u,m}$'s, $\forall u,m$, because each range is associated with at most one target and these assigned ranges should not be considered at the next step. As mentioned above, the complexity will be further reduced at the $(t+1)$-th step, because each $\mathcal{D}_{u,m}$ has a smaller number of ranges such that the size of the feasible region $\tilde{\mathcal{G}}^{(M_{t+1})}$ becomes smaller. Then, we update $t=t+1$ to go into the next step. By repeating the above procedure until $t>M-2$, we can obtain the estimated number of targets and the mapping solution for problem (P2). In the end, with the estimated mapping solutions for all targets, the location of each target can be estimated by solving problem (P1).

\begin{algorithm}[t]
{\bf Input}: $(a_m,b_m)$'s, $\mathcal{D}_{u,m}$'s, $\forall u,m$ \\
{\bf Initialize}: $t=1$, $\hat{K}=0$ and $\hat{\mathcal{G}}=\{ \}$;\\
{\bf Repeat}:
\begin{enumerate}
\item[1.] Apply Algorithm \ref{alg:1} to solve problem (P3) and obtain the solution $\hat{K}_{M_t}$ and $\hat{\mathcal{G}}_1^{(M_t)},\ldots,\hat{\mathcal{G}}_{K_{M_t}}^{(M_t)}$;
\item[2.] Remove $\mathcal{D}_{u,m}(\hat{g}_{u,m,k})$ from $\mathcal{D}_{u,m}$, $\forall \hat{g}_{u,m,k} \in \hat{\mathcal{G}}_k^{(M_t)}$ with $\hat{g}_{u,m,k} >0$, $k=1,\ldots,\hat{K}_{M_t}$, $\forall u,m$;
\item[3.] Infer the mapping solution for the $\hat{K}_{M_t}$ new detected targets, denoted by $\hat{\mathcal{G}}_{\hat{K}+1},\ldots,\hat{\mathcal{G}}_{\hat{K}+K_{M_t}}$
\item[4.] Update $\hat{\mathcal{G}}=\hat{\mathcal{G}} \cup \{\hat{\mathcal{G}}_{\hat{K}+1},\ldots,\hat{\mathcal{G}}_{\hat{K}+K_{M_t}}\}$, $\hat{K}=\hat{K}+\hat{K}_{M_t}$, and $t=t+1$;
\end{enumerate}
{\bf Until} $t>M-2$. \\
{\bf Output}:

1) Obtain the estimated number of targets and the solution of LOS identification and data association, respectively, i.e., $\hat{K}$ and $\hat{\mathcal{G}}=\{ \hat{\mathcal{G}}_k\}_{k=1}^{\hat{K}}$;\\
2) Estimate the target locations by solving problem ({\rm P1}) given the mapping solution $\hat{\mathcal{G}}_k$'s, denoted by $(\hat{x}_k,\hat{y}_k)$'s, $k=1,\ldots,\hat{K}$.

\caption{Algorithm for estimating the mapping and location solutions of all targets}
\label{alg:2}
\end{algorithm}

The details of the above process are shown in Algorithm \ref{alg:2}. Note that solving problem (P2) is decoupled into sequentially finding the solution to problem (P3$-l$). Thus, Algorithm \ref{alg:1} can be applied to each step of Algorithm \ref{alg:2}. Moreover, if there are multiple mapping solutions at the $t$-th step, the one minimizing the localization residual is selected as the optimal solution. After finding the mapping solution for the targets that are exactly detected by $M-t+1$ BSs at the $t$-th step, the corresponding ranges are removed from $\mathcal{D}_{u,m}$'s and should not be considered at the next step.

The complexity of solving problem (P3-$l$) depends on the times of solving problem (P1), because we need to check the constraint (\ref{eqS4.6}) for each mapping $\mathcal{G}_k^{(l)}$, $k=1,\ldots,K_l$. If the constraints (\ref{eqS4.4}) and (\ref{eqS4.5}) are not utilized to filter those ineffective mappings, we may solve problem (P1) many times to check whether (\ref{eqS4.6}) holds for each mapping element in $\mathcal{G}^{(l)}$, whose cardinality is much larger than that of $\bar{\mathcal{G}}^{(l)}$. On the other hand, the complexity of problem (P4) is derermined by the cardinality of $\tilde{\mathcal{G}}^{(l)}$, which is generally small after applying constrains (\ref{eqS4.4})-(\ref{eqS4.6}). In Section \ref{sec:numerical results}, we provide numerical results about the cardinalities of $\mathcal{G}^{(l)}$, $\bar{\mathcal{G}}^{(l)}$, and $\tilde{\mathcal{G}}^{(l)}$ to prove the reduction of computational complexity. Moreover, the complexity of the proposed algorithm can be further reduced by transforming problem (P4) into problem (P5), since the data association possibilities that any two mappings belonging to the same $\tilde{\mathcal{G}}_i^{(l)}$ will not be considered in problem (P5). Denote the cardinality of $\tilde{\mathcal{G}}_i^{(l)}$ by $N_i^{(l)}$ in (\ref{eqS5.6}). Specifically, we just need to try at most $\prod_{i=0}^{N_{1,1}} N_i^{(l)}$ data association possibilities to check whether $(\ref{eqS4.3})$ holds in problem (P5). In exhaustive search method to problem (P4), the number of data association possibilities in problem (P4) is $\tbinom{N_{\tilde{\mathcal{G}}^{(l)}}}{K_l}$ for a given $K_l$ such that the total $\sum_{n=K_l}^{N_{\tilde{\mathcal{G}}^{(l)}}} \tbinom{N_{\tilde{\mathcal{G}}^{(l)}}}{n}$ data association possibilities need to be numerated. 

\begin{remark}
It is worth noting that there may be error propagation in the proposed Algorithm \ref{alg:2}. Define the error propagation event as that before the last step, i.e., $M_t>3$, a spurious mapping $\mathcal{G}_{k_t}^{(M_t)}$ occurs, which satisfies
\begin{align}
&|\{ g_{m,m,k}|g_{m,m,k}>0, g_{m,m,k} \in \mathcal{G}_{k},g_{m,m,k} \in \mathcal{G}_{k_t}^{(M_t)} \}| \nonumber \\
\leq & 2,~\forall k.
\end{align}
In other words, the spurious target $k_t$ with the mapping $\mathcal{G}_{k_t}^{(M_t)}$ shares at most two same direct ranges with any true target $k$. The error propagation event is most likely to happen when $M_t=4$, because there will be more constraints arising from (\ref{eqS4.5}) to prevent errors when $M_t$ is larger. Then, there will be two cases for the target $k_t$ when $M_t=4$: (i) the target $k_t$ shares at most one direct range with any true target; (ii) the target $k_t$ shares two direct ranges with some true target $k$. Case (i) indicates that the four direct ranges are independent of each other and should satisfy (\ref{eqS4.6}) to lead to one target. Besides, there should exist sum ranges that satisfy (\ref{eqS4.4}) and (\ref{eqS4.5}) with these direct ranges. As for Case (ii), the two direct ranges corresponding to the true target $k$ give rise to two possible locations with one being the location of the true target $k$ and the other one being that of the target $k_t$. In this case, the other two direct ranges should be matched with the location of the target $k_t$. Similarly, there should also exist sum ranges that can be associated with the target $k_t$. Therefore, the probability that Case (i) or Case (ii) happens in our proposed algorithm is quite low. In summary, the error propagation will not significantly degrade the performance of target localization.
\end{remark}

\section{Numerical Results}\label{sec:numerical results}
In this section, we provide the numerical results to verify the effectiveness of our proposed two-phase protocol for networked device-free sensing. In the network, we consider $M=4$ BSs and $2 \leq K \leq 7$ targets over a $80$ m $\times$ $80$ m square. To account for LOS blockage, we denote $P_{b}$ by the probability that the LOS path between a target and a BS is blocked. As for Type III paths, we assume that the number of Type III paths for each target between any two BSs follows a Binomial distribution. In detail, the probabilities that there is/is not one Type III path for each target are $P_{\rm nl}$ and $1-P_{\rm nl}$, respectively. Besides, the range of the Type III path is randomly generated in our setup.

\subsection{Joint STO and Range Estimation in Phase I}
\begin{figure}[t]
\centering
\includegraphics[scale=0.56]{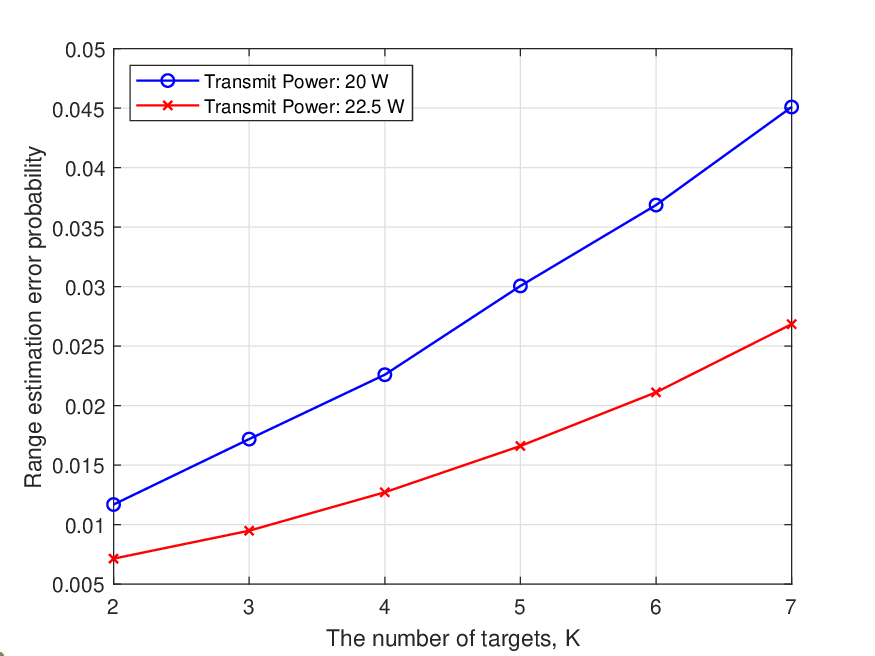}
\caption{Range estimation error probability versus the number of targets.}
\label{range estimation}
\end{figure}

In this part, we provide a numerical example to evaluate the range estimation performance of our proposed algorithm in Phase I. Specifically, we set $N = 3300$ and $\Delta f = 120$ kHz such that $B = 400$ MHz \cite{3gpp}. According to \cite{zaidi2017nr}, with $\Delta f = 120$ kHz, the length of the CP is 0.59 $\mu$s. Besides, the STOs between any two BSs are randomly generated in the interval $[-\tau_{\rm max},\tau_{\rm max}]$, where $\tau_{\rm max}$ is set as 10. To make $L+\tau_{\rm max}<Q$ such that all the ISI is received within the CP, we assume that the maximum number of resolvable paths is $L=200$. Moreover, $P_{\rm b}$ and $P_{\rm nl}$ are set as 0.1 and 0.5, respectively. Under this setup, we randomly generate $10^5$ independent location realizations of BSs and targets, which are uniformly distributed in the considered area. Given the coordinates of BSs and targets, we can know the delays in terms of OFDM sample periods for Type I and Type II paths. Define $\mathcal{L}_{u,m}$ as the set consisting of the indices of the non-zero channel coefficients in ${\boldsymbol{h}}_{u,m}$, $\forall u,m$. Then, we estimate the channels $\tilde{\boldsymbol{h}}_{u,m}$'s by solving problem (\ref{eqS3.2}), and define the set consisting of the indices of the non-zero channel coefficients in $\bar{\boldsymbol{h}}_{u,m}$ as $\bar{\mathcal{L}}_{u,m}$. Based on the estimated channels and the true range between any two BSs, we can estimate the STOs between any two BSs based on (\ref{eqS3.4}) and (\ref{eqS3.5}). With the estimated STO $\bar{\tau}_{u,m}$'s, we will substract each element in $\bar{\mathcal{L}}_{u,m}$ by $\bar{\tau}_{u,m}$ to get $\hat{\mathcal{L}}_{u,m}$, $\forall u,m$., i.e., STO compensation in (\ref{eqS3.6}). If there exist $u$ and $m$ such that $\mathcal{L}_{u,m} \neq \hat{\mathcal{L}}_{u,m}$, we say that the range estimation is in error in this realization. The range estimation error probability versus the number of targets, i.e., $K$, is shown in Fig. \ref{range estimation}, where the BS transmit power is set as 20 Watt (W) and 22.5 W, respectively. It is observed that the range estimation error probability is very low under our proposed scheme, and can be significantly reduced by increasing the transmit power.

\subsection{Localization Accuracy of Two-Phase Protocol}
In this part, we provide numerical examples to verify the effectiveness of the overall two-phase protocol for target localization. To show the effectiveness of our proposed algorithm, we adopt the following schemes as benchmark schemes for performance comparison.
\begin{itemize}
\item \emph{Benchmark Scheme 1:} Under this benchmark scheme, range estimation in Phase I is the same as our method in Section \ref{sec:Phase I}, while in Phase II, there is no constraint (\ref{eqS4.3}) for multiple target localiztion\cite{shen2014estimating}. In other words, each range can be associated with multiple targets. Therefore, each mapping that satisfies the contraints (\ref{eqS4.2}), and (\ref{eqS4.4})-(\ref{eqS4.6}) is used for target localization. Specifically, given each mapping element in $\tilde{\mathcal{G}}^{(l)}$, $\forall l$, problem (P1) is solved for estimating the location of the target corresponding to the mapping.
\item \emph{Benchmark Scheme 2:} Under this benchmark scheme, range estimation in Phase I is the same as our method in Section \ref{sec:Phase I}, while in Phase II, we assume that data association is perfectly known, and we just need to perform LOS identification to localize the targets. As a result, this scheme can serve as an error probability lower bound. With the known data association solution, given each target, each BS knows which ranges estimated in Phase I belong to this range. For these ranges belonging to some target, the job of each BS is to identify the ones of Type I/Type II paths if it has the LOS path to this target. We adopt the LOS identification scheme proposed in \cite{chen1999non} to tackle the above challenge.
\end{itemize}

To compare the performance of the proposed scheme and benchmark schemes, we generate $10^4$ independent location realizations of BSs and targets, which are distributed in the considered area. The setup is same as that introduced in the last sub-section, i.e., $\tau_{\rm max}=10$, $L=200$, $B=400$ MHz, $P_{\rm nl}=0.5$, and $P_{\rm b}=0.1$. Besides, the transmit power is set as 20 W. In each realization, we first estimate channels by solving problem (10) to obtain $\mathcal{D}_{u,m}$'s based on (\ref{eqS3.8}), and then estimate the number and the locations of targets by Algorithm \ref{alg:2}, Benchmark 1, and Benchmark 2, respectively. A target is said to be localized correctly if the distance between its estimated location and its true location is no larger than $r$ m. Denote $N_i$ and $K_i$ as the numbers of the correctly localized targets and the detected targets in the $i$-th realization, respectively. Then, miss detection (MD) and false alarm (FA) error probabilities over the $10^4$ realizations, which are defined as $P_{\rm MD}=\frac{\sum_{i=1}^{10^4}(K-N_i)}{K \times 10^4}$ and $P_{\rm FA}=\frac{\sum_{i=1}^{10^4}(K_i-N_i)}{K \times 10^4}$, respectively, are used to characterize the target localization performance for each method.

\begin{figure}[t]
\centering
\subfigure[Miss detection probability $P_{\rm MD}$ ]
{\scalebox{0.56}{\includegraphics*{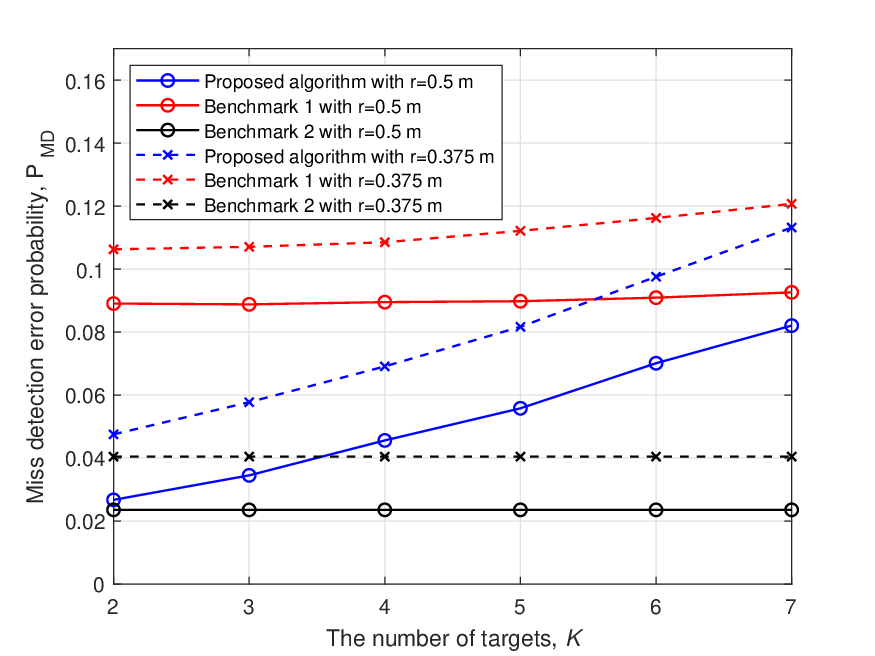}}}
\subfigure[False alarm probability $P_{\rm FA}$ ]
{\scalebox{0.56}{\includegraphics*{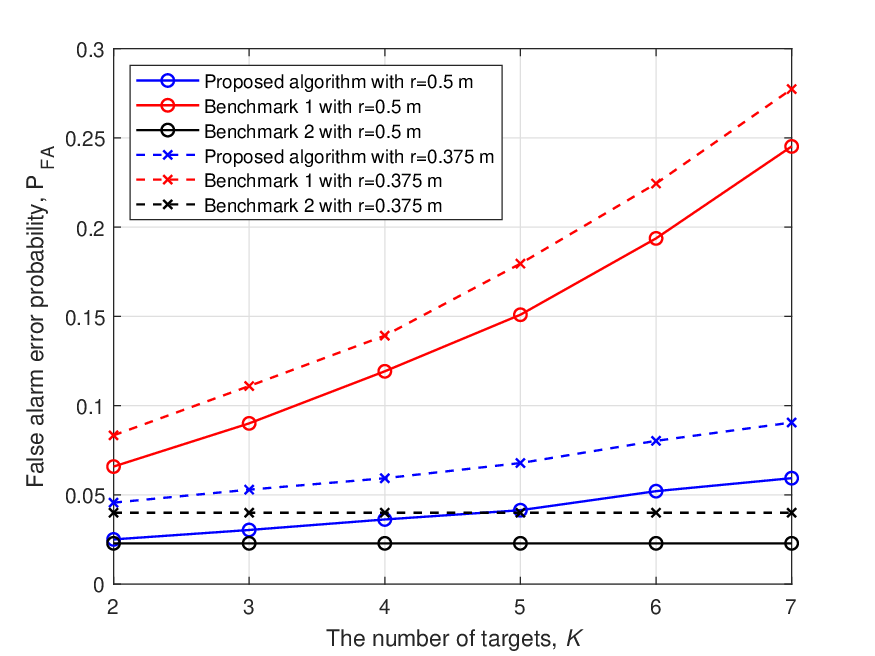}}}
\caption{The comparison of location estimation error probability between the benchmark and the proposed algorithm with different $r$ when $B=400$ MHz, $P_{\rm b}=0.1$, and $P_{\rm nl}=0.5$.}
\label{fig3}
\end{figure}
Fig. \ref{fig3} gives the comparison between the benchmarks and the proposed algorithm in terms of $P_{\rm MD}$ and $P_{\rm FA}$ with different values of $r$. It is observed that under our proposed scheme, the probabilities of miss detection and false alarm are below $12\%$ and $9\%$ when $K$ ranges from 2 to 7 and with $r=0.375$ m and $r=0.5$ m, respectively. We can also see that the proposed algorithm outperforms Benchmark 1 in terms of both miss detection and false alarm error probabilities when $K$ ranges from 2 to 7. Moreover, there is merely a small performance gap between the proposed algorithm and Benchmark 2, where data association is assumed to be known. Therefore, the performance of our proposed joint data association and LOS identification scheme is very close to the localization error probability lower bound.

\begin{figure}[t]
\centering
\includegraphics[scale=0.56]{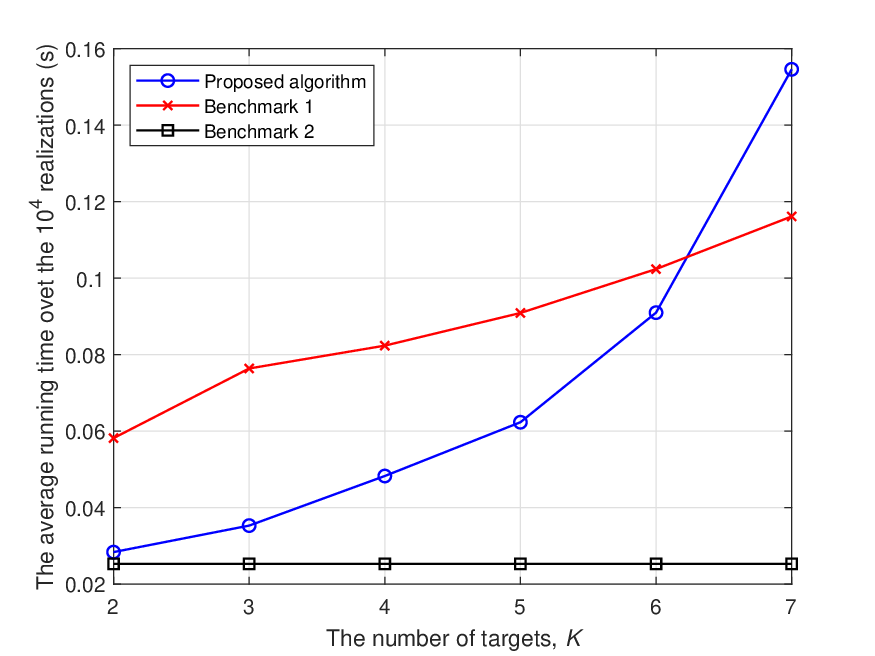}
\caption{The comparison of the average running time for the benchmarks and the proposed algorithm when $B=400$ MHz, $P_{\rm b}=0.1$, and $P_{\rm nl}=0.5$.}
\label{fig4}
\end{figure}

As for the complexity, the three methods are compared in terms of the average running time in seconds (s) over the $10^4$ realizations, which is depicted in Fig. \ref{fig4}. We can see that the average running time for the proposed scheme is less than 0.16 seconds in each realization when $K$ ranges from 2 to 7. Considering each BS will have much more computational resources, the actual running time will be lower. Besides, the running time for Benchmark 1 is longer than that of our proposed algorithm when the number of targets is small. There are two reasons, (i) since all possible mappings satisfying constraints (\ref{eqS4.2}), and (\ref{eqS4.4})-(\ref{eqS4.6}) will be retained, the times of solving problem (P1) in Benchmark 1 is more than that in the proposed algorithm; (ii) the complexity of solving problem (P5) will not be high in the proposed algorithm when the number of targets is small. However, when the number of targets is larger, e.g., $K=7$, the complexity of solving problem (P5) will be higher such that the running time for the proposed algorithm is longer.

\begin{figure}[t]
\centering
\includegraphics[scale=0.56]{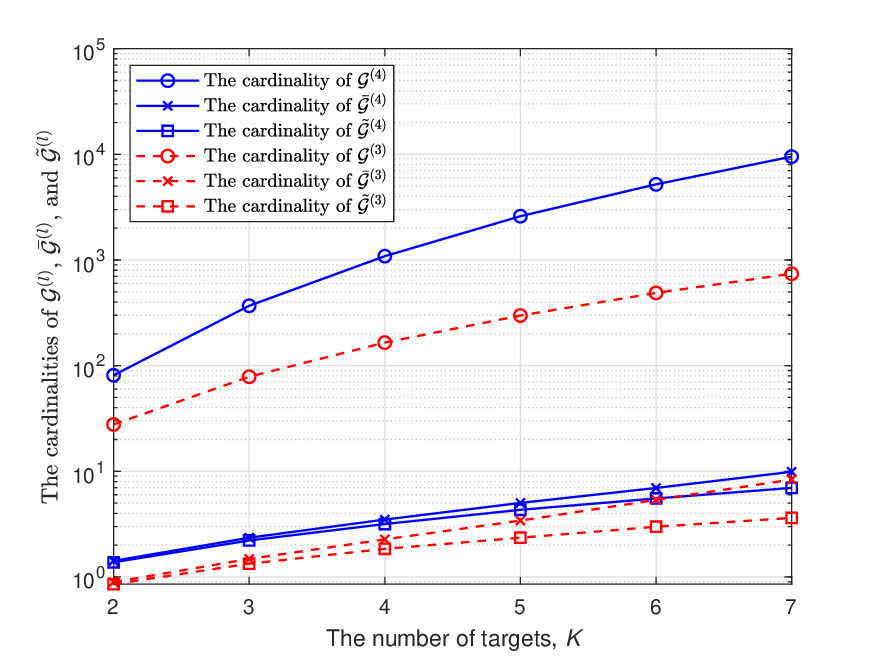}
\caption{The cardinalities of $\mathcal{G}^{(l)}$, $\bar{\mathcal{G}}^{(l)}$, and $\tilde{\mathcal{G}}^{(l)}$ for the proposed algorithm when $B=400$ MHz, $P_{\rm b}=0.1$, and $P_{\rm nl}=0.5$.}
\label{fig5}
\end{figure}

On the other hand, we also provide the simulation results for the average cardinalities of $\mathcal{G}^{(l)}$, $\bar{\mathcal{G}}^{(l)}$, and $\tilde{\mathcal{G}}^{(l)}$ over the $10^4$ realizations, as shown in Fig. \ref{fig5}. It is observed that via utilizing (\ref{eqS4.4}) and (\ref{eqS4.5}), the cardinality of $\bar{\mathcal{G}}^{(l)}$ is much smaller than that of $\mathcal{G}^{(l)}$, $\forall l \in \{3,4\}$. In other words, the sum distance constraints can efficiently remove the bad data association solutions in $\mathcal{G}^{(l)}$, $\forall l \in \{3,4\}$. Therefore, thanks to constraints (\ref{eqS4.4})-(\ref{eqS4.6}), the utilizations of solving problem (P1) can be greatly reduced. Moreover, the number of possible mapping solutions can be further reduced by (\ref{eqS4.6}), i.e., the cardinality of $\tilde{\mathcal{G}}^{(l)}$ is fewer than that of $\bar{\mathcal{G}}^{(l)}$, $\forall l \in \{3,4\}$, which can reduce the complexity for solving problem (P4).

\begin{figure}[t]
\centering
\subfigure[Miss detection probability $P_{\rm MD}$ ]
{\scalebox{0.56}{\includegraphics*{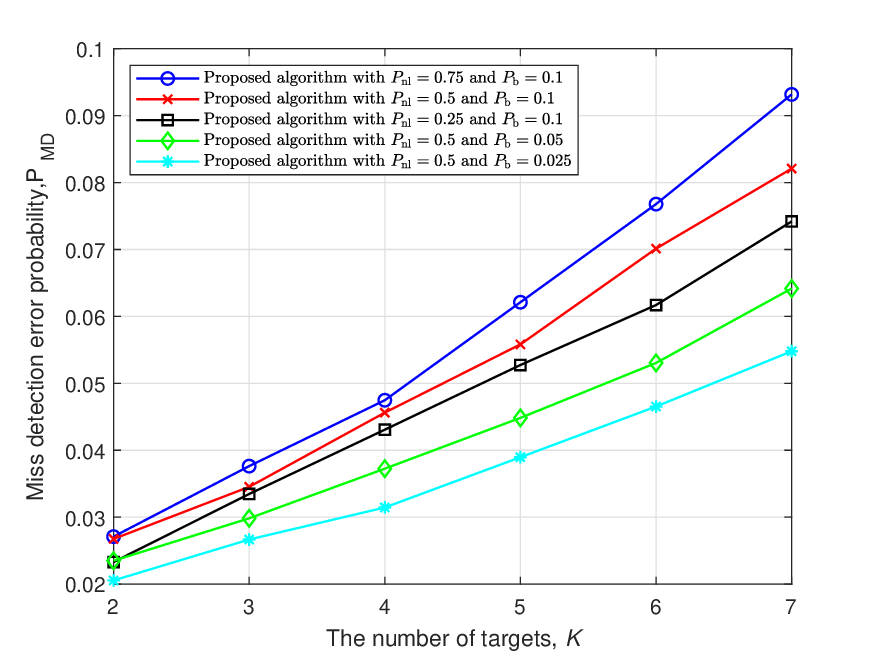}}}
\subfigure[False alarm probability $P_{\rm FA}$ ]
{\scalebox{0.56}{\includegraphics*{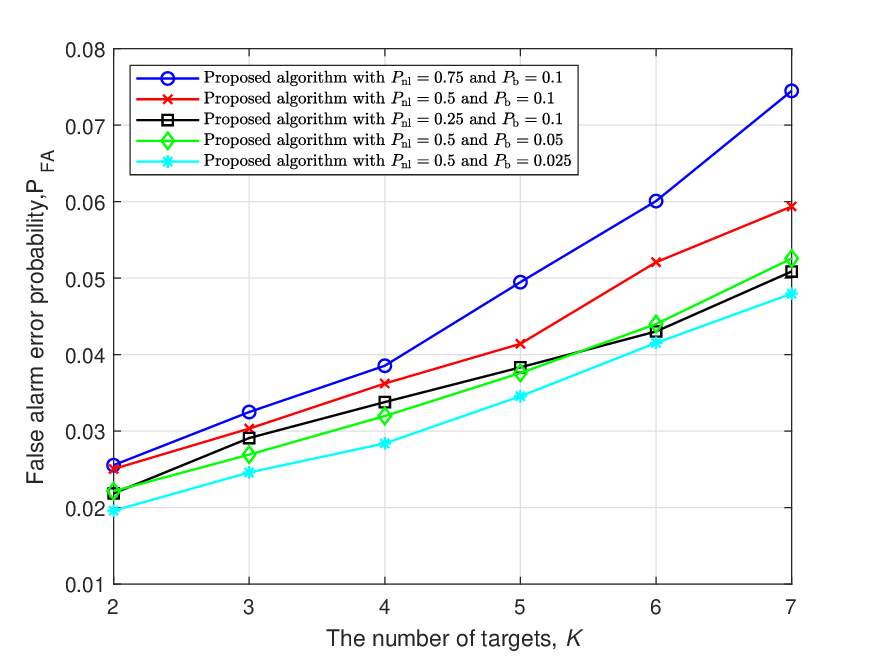}}}
\caption{The comparison of location estimation error probability between the benchmark and the proposed algorithm with different $P_{\rm b}$ and $P_{\rm nl}$ when $B=400$ MHz and $r=0.5$ m.}
\label{fig6}
\end{figure}

\begin{figure}[t]
\centering
\subfigure[Miss detection probability $P_{\rm MD}$ ]
{\scalebox{0.56}{\includegraphics*{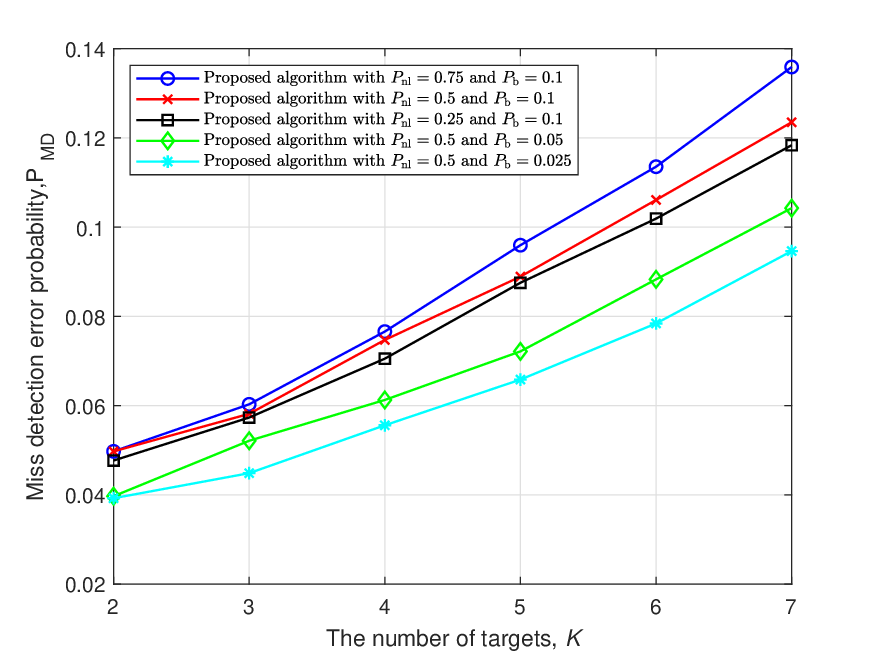}}}
\subfigure[False alarm probability $P_{\rm FA}$ ]
{\scalebox{0.56}{\includegraphics*{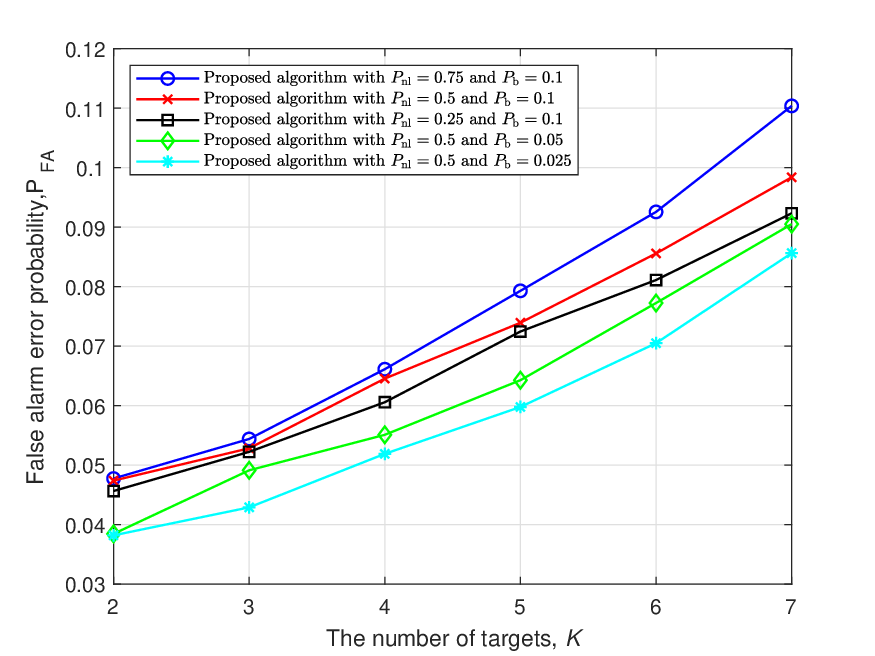}}}
\caption{The comparison of location estimation error probability between the benchmark and the proposed algorithm with different $P_{\rm b}$ and $P_{\rm nl}$ when $B=300$ MHz and $r=0.5$ m.}
\label{fig7}
\end{figure}

\subsection{Effect of LOS Blockage, NLOS Paths, and Bandwidth on Performance}
In this sub-section, we show the impact of LOS blockage, NLOS paths, and bandwidth on target localization. Specifically, the probability that determines the number of Type III paths, i.e., $P_{\rm nl}$, is set as 0.75, 0.5, and 0.25, respectively. The bandwidth that determines the range resolution, i.e., $B$, is set as 400 MHz and 300 MHz, respectively. The probability that the LOS path between a target and a BS is blocked, i.e., $P_{\rm b}$, is set as 0.1, 0.05, and 0.025, respectively.

The effect of bandwidth, LOS blockage, and NLOS paths on target localization
for the proposed method is shown in Fig. \ref{fig6} and Fig. \ref{fig7} with $r=0.5$ m. It is depicted that the location error probability will be lower when the probability $P_{\rm b}$ is decreased, since the additional LOS path information is beneficial for target localization. Moreover, we can see that when the probability of the existence of NLOS paths $P_{\rm nl}$ is increased, the location error probabilities of our proposed algorithm increase as well. The reason is that the NLOS paths will increase the probability of the existence of spurious mappings such that the localization performance is degraded. It is observed from Fig. \ref{fig6} and Fig. \ref{fig7} that the target localization performance can be improved by increasing the system bandwidth from 300 MHz to 400 MHz. Because the larger bandwidth can bring a higher range resolution, the accuracy of target localization in our proposed algorithm will be enhanced.

\section{Conclusion}\label{sec:conclusion}
In this paper, we study networked device-free sensing based on the echoes of the transmitted downlink communication signals in a multi-path environment, where the BSs are not perfectly synchronized but will share the communication and sensing information. A two-phase protocol is adopted. In the first phase, each BS estimates the direct distances from targets to itself and the sum distances from other BSs to targets to itself. Since the range of the LOS path between any two BSs is known, it is utilized to estimate the STOs such that the range of each path between any two BSs can be accurately estimated. In the second phase, all BSs will collaboratively localize targets based on range estimations obtained in Phase I. Nevertheless, there are issues of range interference arising from NLOS paths, LOS blockage for some BSs and some targets, and the unknown mapping between ranges of LOS paths and targets, which are coupled together. To tackle the above challenges, an efficient algorithm is proposed to iteratively localize the targets that are exactly detected by a given number of BSs. In this way, we can decouple the issue of LOS blockage from the problem of NLOS paths and data association. As a result, the number and the locations of targets can be efficiently estimated. Numerical results show that the proposed strategy can achieve high accuracy of target localization, thanks to BS coordination and large bandwidth in the 6G cellular network.

\bibliographystyle{IEEEtran}
\bibliography{ref}

\begin{thebibliography}{10}
\providecommand{\url}[1]{#1}
\csname url@samestyle\endcsname
\providecommand{\newblock}{\relax}
\providecommand{\bibinfo}[2]{#2}
\providecommand{\BIBentrySTDinterwordspacing}{\spaceskip=0pt\relax}
\providecommand{\BIBentryALTinterwordstretchfactor}{4}
\providecommand{\BIBentryALTinterwordspacing}{\spaceskip=\fontdimen2\font plus
\BIBentryALTinterwordstretchfactor\fontdimen3\font minus
  \fontdimen4\font\relax}
\providecommand{\BIBforeignlanguage}[2]{{%
\expandafter\ifx\csname l@#1\endcsname\relax
\typeout{** WARNING: IEEEtran.bst: No hyphenation pattern has been}%
\typeout{** loaded for the language `#1'. Using the pattern for}%
\typeout{** the default language instead.}%
\else
\language=\csname l@#1\endcsname
\fi
#2}}
\providecommand{\BIBdecl}{\relax}
\BIBdecl

\bibitem{shi2023joint}
Q.~Shi, L.~Liu, and S.~Zhang, ``Joint data association, {NLOS} mitigation, and
  clutter suppression for networked device-free sensing in {6G} cellular
  network,'' in \emph{Proc. IEEE Int. Conf. Acoust., Speech and Signal Process.
  (ICASSP)}, 2023, pp. 1--5.

\bibitem{IMT}
ITU-R, ``M.2160 : Framework and overall objectives of the future development of
  imt for 2030 and beyond,'' \emph{[Online]. Available:
  https://www.itu.int/rec/R-REC-M.2160/en}, 2023.

\bibitem{mu2022noma}
X.~Mu, Y.~Liu, L.~Guo, J.~Lin, and L.~Hanzo, ``{NOMA-aided joint radar and
  multicast-unicast communication systems},'' \emph{IEEE J. Sel. Areas
  Commun.}, vol.~40, no.~6, pp. 1978--1992, Jun. 2022.

\bibitem{tsinos2021joint}
C.~G. Tsinos, A.~Arora, S.~Chatzinotas, and B.~Ottersten, ``Joint transmit
  waveform and receive filter design for dual-function radar-communication
  systems,'' \emph{IEEE J. Sel. Topics Signal Process.}, vol.~15, no.~6, pp.
  1378--1392, Nov. 2021.

\bibitem{liu2020joint}
X.~Liu, T.~Huang, N.~Shlezinger, Y.~Liu, J.~Zhou, and Y.~C. Eldar, ``Joint
  transmit beamforming for multiuser {MIMO} communications and {MIMO} radar,''
  \emph{IEEE Trans. Signal Process.}, vol.~68, pp. 3929--3944, Jun. 2020.

\bibitem{sturm2011waveform}
C.~Sturm and W.~Wiesbeck, ``Waveform design and signal processing aspects for
  fusion of wireless communications and radar sensing,'' \emph{Proc. IEEE},
  vol.~99, no.~7, pp. 1236--1259, Jul. 2011.

\bibitem{zheng2017super}
L.~Zheng and X.~Wang, ``{Super-resolution delay-doppler estimation for OFDM
  passive radar},'' \emph{IEEE Trans. Signal Process.}, vol.~65, no.~9, pp.
  2197--2210, May 2017.

\bibitem{liu2020two}
L.~Liu and S.~Zhang, ``{A two-stage radar sensing approach based on MIMO-OFDM
  technology},'' in \emph{Proc. IEEE Globecom Workshops (GC) Wkshps}.\hskip 1em
  plus 0.5em minus 0.4em\relax IEEE, 2020, pp. 1--6.

\bibitem{gaudio2020effectiveness}
L.~Gaudio, M.~Kobayashi, G.~Caire, and G.~Colavolpe, ``{On the effectiveness of
  OTFS for joint radar parameter estimation and communication},'' \emph{IEEE
  Trans. Wireless Commun.}, vol.~19, no.~9, pp. 5951--5965, Sep. 2020.

\bibitem{dokhanchi2019mmwave}
S.~H. Dokhanchi, B.~S. Mysore, K.~V. Mishra, and B.~Ottersten, ``{A mmWave
  automotive joint radar-communications system},'' \emph{IEEE Trans. Aerosp.
  Electron. Syst.}, vol.~55, no.~3, pp. 1241--1260, Jun. 2019.

\bibitem{barneto2022millimeter}
C.~B. Barneto, E.~Rastorgueva-Foi, M.~F. Keskin, T.~Riihonen, M.~Turunen,
  J.~Talvitie, H.~Wymeersch, and M.~Valkama, ``{Millimeter-wave mobile sensing
  and environment mapping: Models, algorithms and validation},'' \emph{IEEE
  Trans. Veh. Technol.}, vol.~71, no.~4, pp. 3900--3916, Apr. 2022.

\bibitem{yang2022hybrid}
J.~Yang, C.-K. Wen, and S.~Jin, ``{Hybrid active and passive sensing for SLAM
  in wireless communication systems},'' \emph{IEEE J. Sel. Areas Commun.}, Jul.
  2022.

\bibitem{liu2015joint}
L.~Liu, S.~Bi, and R.~Zhang, ``Joint power control and fronthaul rate
  allocation for throughput maximization in ofdma-based cloud radio access
  network,'' \emph{IEEE Trans. Commun.}, vol.~63, no.~11, pp. 4097--4110, Nov.
  2015.

\bibitem{gesbert2010multi}
D.~Gesbert, S.~Hanly, H.~Huang, S.~S. Shitz, O.~Simeone, and W.~Yu,
  ``Multi-cell mimo cooperative networks: A new look at interference,''
  \emph{IEEE J. Sel. Areas Commun.}, vol.~28, no.~9, pp. 1380--1408, Dec. 2010.

\bibitem{checko2014cloud}
A.~Checko, H.~L. Christiansen, Y.~Yan, L.~Scolari, G.~Kardaras, M.~S. Berger,
  and L.~Dittmann, ``Cloud ran for mobile networks—a technology overview,''
  \emph{IEEE Commun. Surveys Tuts.}, vol.~17, no.~1, pp. 405--426, Firstquarter
  2014.

\bibitem{liu2022survey}
A.~Liu, Z.~Huang, M.~Li, Y.~Wan, W.~Li, T.~X. Han, C.~Liu, R.~Du, D.~K.~P. Tan,
  J.~Lu \emph{et~al.}, ``A survey on fundamental limits of integrated sensing
  and communication,'' \emph{IEEE Commun. Surveys Tuts.}, vol.~24, no.~2, pp.
  994--1034, Feb. 2022.

\bibitem{xie2023collaborative}
L.~Xie, S.~Song, Y.~C. Eldar, and K.~B. Letaief, ``Collaborative sensing in
  perceptive mobile networks: Opportunities and challenges,'' \emph{IEEE
  Wireless Commun.}, vol.~30, no.~1, pp. 16--23, Feb. 2023.

\bibitem{zhang2020perceptive}
A.~Zhang, M.~L. Rahman, X.~Huang, Y.~J. Guo, S.~Chen, and R.~W. Heath,
  ``Perceptive mobile networks: Cellular networks with radio vision via joint
  communication and radar sensing,'' \emph{IEEE Veh. Technol. Mag.}, vol.~16,
  no.~2, pp. 20--30, Jun. 2020.

\bibitem{mao2007wireless}
G.~Mao, B.~Fidan, and B.~D. Anderson, ``Wireless sensor network localization
  techniques,'' \emph{Comput. Netw.}, vol.~51, no.~10, pp. 2529--2553, Jan.
  2007.

\bibitem{aditya2018survey}
S.~Aditya, A.~F. Molisch, and H.~M. Behairy, ``A survey on the impact of
  multipath on wideband time-of-arrival based localization,'' \emph{Proc.
  IEEE}, vol. 106, no.~7, pp. 1183--1203, Jul. 2018.

\bibitem{guvenc2009survey}
I.~Guvenc and C.-C. Chong, ``{A survey on TOA based wireless localization and
  NLOS mitigation techniques},'' \emph{IEEE Commun. Surveys Tuts.}, vol.~11,
  no.~3, pp. 107--124, Aug. 2009.

\bibitem{chen2013comparative}
X.~Chen, F.~Dovis, S.~Peng, and Y.~Morton, ``{Comparative studies of GPS
  multipath mitigation methods performance},'' \emph{IEEE Trans. on Aerosp.
  Electron. Syst.}, vol.~49, no.~3, pp. 1555--1568, Jul. 2013.

\bibitem{kotaru2015spotfi}
M.~Kotaru, K.~Joshi, D.~Bharadia, and S.~Katti, ``{Spotfi: Decimeter level
  localization using wifi},'' in \emph{Proc. ACM SIGCOMM Comput. Commun. Rev.},
  vol.~45, 2015, pp. 269--282.

\bibitem{chen1999non}
P.-C. Chen, ``A non-line-of-sight error mitigation algorithm in location
  estimation,'' in \emph{Proc. IEEE Int. Conf. Wireless Commun. Networking
  (WCNC)}, vol.~1, Sep. 1999, pp. 316--320.

\bibitem{venkatesh2006linear}
S.~Venkatesh and R.~M. Buehrer, ``{A linear programming approach to NLOS error
  mitigation in sensor networks},'' in \emph{Proc. IEEE Int. Symp. Inf.
  Processing Sensor Netw. (IPSN)}, Apr. 2006, pp. 301--308.

\bibitem{decarli2010nlos}
N.~Decarli, D.~Dardari, S.~Gezici, and A.~A. D'Amico, ``{LOS/NLOS detection for
  UWB signals: A comparative study using experimental data},'' in \emph{Proc.
  5th IEEE Int. Symp. Wireless Pervasive Comput.}, May. 2010, pp. 169--173.

\bibitem{shi2022device}
Q.~Shi, L.~Liu, S.~Zhang, and S.~Cui, ``Device-free sensing in {OFDM} cellular
  network,'' \emph{IEEE J. Sel. Areas Commun.}, vol.~40, no.~6, pp. 1838--1853,
  Jan. 2022.

\bibitem{poore1994multidimensional}
A.~B. Poore, ``Multidimensional assignment formulation of data association
  problems arising from multitarget and multisensor tracking,''
  \emph{Computational Optimization and Applications}, vol.~3, no.~1, pp.
  27--57, Mar. 1994.

\bibitem{poore1993lagrangian}
A.~P. Poore and N.~Rijavec, ``{A Lagrangian relaxation algorithm for
  multidimensional assignment problems arising from multitarget tracking},''
  \emph{SIAM J. Optim.}, vol.~3, no.~3, pp. 544--563, 1993.

\bibitem{kuhn1955hungarian}
H.~W. Kuhn, ``The hungarian method for the assignment problem,'' \emph{Nav.
  Res. Logist. Quart.}, vol.~2, no.~1, pp. 83--97, 1955.

\bibitem{bar1975tracking}
Y.~Bar-Shalom and E.~Tse, ``Tracking in a cluttered environment with
  probabilistic data association,'' \emph{Automatica}, vol.~11, no.~5, pp.
  451--460, Jan. 1975.

\bibitem{fortmann1983sonar}
T.~Fortmann, Y.~Bar-Shalom, and M.~Scheffe, ``Sonar tracking of multiple
  targets using joint probabilistic data association,'' \emph{IEEE J. Ocean.
  Eng.}, vol.~8, no.~3, pp. 173--184, Jul. 1983.

\bibitem{reid1979algorithm}
D.~Reid, ``An algorithm for tracking multiple targets,'' \emph{IEEE Trans.
  Automat. Contr.}, vol.~24, no.~6, pp. 843--854, Dec. 1979.

\bibitem{shen2014estimating}
J.~Shen and A.~F. Molisch, ``Estimating multiple target locations in multi-path
  environments,'' \emph{IEEE Trans. Wireless Commun.}, vol.~13, no.~8, pp.
  4547--4559, Aug. 2014.

\bibitem{schmidl1997robust}
T.~M. Schmidl and D.~C. Cox, ``Robust frequency and timing synchronization for
  {OFDM},'' \emph{IEEE Trans. Commun.}, vol.~45, no.~12, pp. 1613--1621, Dec.
  1997.

\bibitem{morelli2007synchronization}
M.~Morelli, C.-C.~J. Kuo, and M.-O. Pun, ``Synchronization techniques for
  orthogonal frequency division multiple access ({OFDMA}): A tutorial review,''
  \emph{Proc. of IEEE}, vol.~95, no.~7, pp. 1394--1427, July 2007.

\bibitem{LASSO}
M.~Yuan and Y.~Lin, ``Model selection and estimation in regression with grouped
  variables,'' \emph{J. R. Stat. Soc. Ser. B, Stat. Methodol.}, vol.~68, no.~1,
  pp. 49--67, Feb. 2006.

\bibitem{li2017analysis}
H.~Li, L.~Han, R.~Duan, and G.~M. Garner, ``{Analysis of the synchronization
  requirements of 5G and corresponding solutions},'' \emph{IEEE Commun. Stand.
  Mag.}, vol.~1, no.~1, pp. 52--58, Mar. 2017.

\bibitem{hwang2008ofdm}
T.~Hwang, C.~Yang, G.~Wu, S.~Li, and G.~Y. Li, ``{OFDM and its wireless
  applications: A survey},'' \emph{IEEE Trans. Veh. Technol.}, vol.~58, no.~4,
  pp. 1673--1694, May 2008.

\bibitem{liu2021uplink}
L.~Liu, Y.-F. Liu, P.~Patil, and W.~Yu, ``Uplink-downlink duality between
  multiple-access and broadcast channels with compressing relays,'' \emph{IEEE
  Trans. Inf. Theory}, vol.~67, no.~11, pp. 7304--7337, Nov. 2021.

\bibitem{grant2009cvx}
M.~Grant, S.~Boyd, and Y.~Ye, ``Cvx: Matlab software for disciplined convex
  programming,'' \emph{http://cvxr.com/cvx}, Mar. 2009.

\bibitem{3gpp}
3GPP, ``{Summary of Release 15 Work Items},'' 3rd Generation Partnership
  Project (3GPP), TR 21.915 Version 1.1.0 Release 15, 2018.

\bibitem{torrieri1984statistical}
D.~J. Torrieri, ``Statistical theory of passive location systems,'' \emph{IEEE
  Trans. Aerosp. Electron. Syst.}, no.~2, pp. 183--198, Mar. 1984.

\bibitem{neira2001data}
J.~Neira and J.~D. Tard{\'o}s, ``Data association in stochastic mapping using
  the joint compatibility test,'' \emph{IEEE Trans. Robot. Autom.}, vol.~17,
  no.~6, pp. 890--897, Dec. 2001.

\bibitem{bailey2000data}
T.~Bailey, E.~M. Nebot, J.~Rosenblatt, and H.~F. Durrant-Whyte, ``{Data
  association for mobile robot navigation: A graph theoretic approach},'' in
  \emph{IEEE Int. Conf. Robot. Automat.}, vol.~3, 2000, pp. 2512--2517.

\bibitem{grimson1991object}
W.~E.~L. Grimson, \emph{Object recognition by computer: the role of geometric
  constraints}.\hskip 1em plus 0.5em minus 0.4em\relax Mit Press, 1991.

\bibitem{zaidi2017nr}
A.~Zaidi \emph{et~al.}, ``{Designing for the future: the 5G NR physical
  layer},'' \emph{Ericsson Technol. Rev.}, 2017.

\end{thebibliography}

\end{document}